\documentclass[preprint,aps,showpacs,nofootinbib,preprintnumbers,amsmath,amssymb]{revtex4-1}
\usepackage{epsfig}
\usepackage{subfigure}
\usepackage{dcolumn}% Align table columns on decimal point
\usepackage{bm}% bold math
\usepackage[usenames ,dvipsnames]{xcolor}
\usepackage{slashed}
\usepackage{graphicx,color}
\usepackage{amsmath}
\usepackage{nonfloat}
\begin{document}
\title{Up-down asymmetries of  charmed baryon three-body decays}
% with SU(3) flavor symmetry}

\author{Jian-Yong Cen$^{1}$, Chao-Qiang Geng$^{2,3,1}$, Chia-Wei Liu$^{2}$ and Tien-Hsueh Tsai$^{2}$}
\affiliation{
	$^{1}$School of Physics and Information Engineering, Shanxi Normal University, Linfen 041004\\
	$^{2}$Department of Physics, National Tsing Hua University, Hsinchu 300\\
	$^{3}$Physics Division, National Center for Theoretical Sciences, Hsinchu 300
}
\date{\today}
\begin{abstract}
We study the up-down asymmetries in the three-body anti-triplet charmed baryon decays of ${\bf B_c}\to{\bf B_n}MM' $ with the $SU(3)_f$ flavor symmetry,  where ${\bf B_c}$ presents the anti-triplet charmed baryon of $(\Xi_c^0,-\Xi_c^+,\Lambda_c^+)$, while ${\bf B_n}$ and $M^{(')}$ denote octet baryon and meson states, respectively. By assuming  the s-wave meson-pairs  to be the dominant constituents  in final state configurations, we can write  the spin-dependent decay amplitude  into  parity-conserving  and parity-violating   parts, parametrized  by 6  real parameters under $SU(3)_f$, respectively. 
 Fitting these parameters by 16 experimental data points with the minimum $\chi^2$ method, we obtain that  $\chi^2/d.o.f=2.4$. With the fitted parameters, we evaluate the up-down asymmetries along with  the decay branching ratios of  ${\bf B_c}\to{\bf B_n}MM' $.  Some of these up-down asymmetries are  accessible to the experiments  at BESIII, BELLE-II and LHCb.
\end{abstract}
\maketitle

\section{introduction}
The three-body charmed baryon decays of ${\bf B_c}\to {\bf B_n} MM'$   
have been recently searched  by the experimental Collaborations of  BELLE, BESIII and LHCb,
where ${\bf B_c}\equiv (\Xi_c^0,-\Xi_c^+,\Lambda_c^+)$ 
denotes the charmed baryon anti-triplet,
while   $\bf B_n$ and $M^{(')}$ correspond to 
the baryon and meson octets, respectively.
In particular, the golden mode of $\Lambda_c^+\to p K^-\pi^+$
has been measured with high precision by BELLE and BESIII~\cite{Zupanc:2013iki,Ablikim:2015flg},
which can improve the accuracies of other $\Lambda_c^+$ decays which are usually given by the rates relative to it~\cite{pdg}.
Since the decay branching ratio is a spin-averaged observable, it loses the ability to probe the polarization property of the parent or daughter baryon in 
	a baryonic decay.  
On the other hand, several P and CP or T violating spin correlations can be constructed  due to the rich spin structures  in the three-body baryonic decays.	
	In order to understand the full dynamics of decay processes, a systematic study, which has been done in two-body modes \cite{Geng:2019xbo}, on both branching ratios and up-down asymmetries, also known as daughter-baryon spin polarizations,  in the three-body charmed baryon modes is necessary. 
	In this work, we only concentrate on the P violating up-down asymmetries
	in ${\bf B_c}\to {\bf B_n} MM'$, which have not been  examined both experimentally and theoretically in the literature yet. 

However, it is known that the investigation into the charm baryon decays has always been difficult.
The main reason for this is  that the scale of the charm quark mass 
%($m_c$) 
is too large for the flavor $SU(4)_f$  symmetry, so that
the  heavy to light quark transitions $(c\to q)$ in charmed decays cannot be easily calculated.
Moreover, the factorization method  fails in these decays~\cite{Bjorken:1988ya}, in addition to that  the three-body processes are much 
more complicated than the two-body ones.
The alternative approaches for the charmed hadron decays 
have been shown in Refs.~~\cite{Cheng:1991sn,Cheng:1993gf,
Zenczykowski:1993hw,Fayyazuddin:1996iy,Dhir:2015tja,Cheng:2018hwl},
where the non-factorizable effects are taken into account.
On the other hand, 
the $SU(3)_f$ symmetry 
has been tested as a useful tool 
in the charmed hadronic decays~\cite{Savage:1989qr,Lu:2016ogy,Pirtskhalava:2011va,Grossman:2012ry,Geng:2017esc,Geng:2018plk,Geng:2017mxn,Geng:2018bow,Geng:2018rse,Geng:2018upx,Geng:2019xbo,Geng:2019bfz,a1,a2,a3}.
%\cite{Savage:1989qr,Sa}

In order to study ${\bf B_c}\to {\bf B_n} MM'$,
we assume that the final state configurations of the meson-pairs are dominated by the s-wave ones, so that  
 the  three-body system can be treated as an effective two-body one with an off-shell scalar meson~\cite{Geng:2018upx} 
to analogize the  three-body  semileptonic decays of charmed baryons~\cite{Geng:2019bfz}. 
Similar to the discussions on the up-down asymmetries in the two-body decays of ${\bf B_c}\to {\bf B_n} M$ in Ref.~\cite{Geng:2019xbo},
we define the spin dependent decay amplitudes  in terms of parity-conserving and violating parts under $SU(3)_f$, respectively,
 resulting in 12 real parameters to be fitted with 16 available data points.
 In our study, we also consider the  the kinematic correction factors as in Ref.~\cite{Geng:2018upx}
to keep the triangle relations derived in Refs.~\cite{Geng:2018upx,Lu:2016ogy,Gronau:2018vei},
but break those by the U-spin symmetry~\cite{Grossman:2018ptn} due to the large differences of hadron masses.

This paper is organized as follows. In  Sec. II, we give the formalism and display  the explicit  amplitudes for the three-body charmed baryon decays 
of  ${\bf B_c}\to{\bf B_n}MM' $ under the $SU(3)_f$ symmetry.
In Sec. III, we present our numerical results and discussions. Our conclusions are shown in Sec. IV.

\section{Formalism}
In order to study the up-down asymmetries of the three-body ${\bf B_c}\to {\bf B_n} MM'$ non-leptonic decays,
we start with the  charm quark decays of $c\to s \bar u d$, $c\to u \bar d d\,(u \bar s s)$ and $c\to d \bar {u} s$
%,where ${\bf B_{c,n}}$ %\equiv (\Xi_c^0,-\Xi_c^+,\Lambda_c^+)$ and $M^{(')}$ denote %being denoted as
%the  octet baryon %anti-triplet  and octet meson states, respectively. 
%
at  tree-level with the effective Hamiltonian, given by~\cite{Buras:1998raa}
\begin{eqnarray}\label{Heff}
{\cal H}_{eff}&=&\sum_{i=-,+}\frac{G_F}{\sqrt 2}c_i
\bigg[V_{cs}V_{ud}^{*} O_i^{ds} +V_{cd}V_{ud}^{*}O_i^{qq} +V_{cd}V_{us}^{*}O_i^{sd}\bigg]\,,
\end{eqnarray}
with
\begin{eqnarray}\label{O12}
	O_\pm^{q_1q_2}&=&{1\over 2}\left[(\bar u q_1)_{V-A}(\bar q_2 c))_{V-A}\pm (\bar q_2 q_1))_{V-A}(\bar u c))_{V-A}\right]\,,\;\nonumber \\
	O_\pm^{qq}&=&O_\pm^{dd}-O_\pm^{ss}\,,\;
\end{eqnarray}
where $G_F$ is the Fermi constant, $c_{\pm}$ represent the Wilson coefficients and
$(V_{cs}V_{ud}$, $V_{cd}V_{ud}$, $V_{cd}V_{us})$
=$c^2_c(1,-t_c,-t_c^2)$ correspond to the CKM matrix elements with $c_c= \cos\theta_c$, 
$t_c=\sin\theta_c/\cos\theta_c$ and $\theta_c$  the Cabibbo angle.
%~\cite{pdg} 
Here, the relation of  $V_{cs}V_{us}=-V_{cd}V_{ud}$ has been used to combine
the  $c\to u d\bar d\,(u s\bar s)$ transitions,
%In Eq.~(\ref{Heff}), 
 $(\bar q_1 q_2)_{V-A}(\bar q_3 c)_{V-A}\equiv
\bar q_1\gamma_\mu(1-\gamma_5)q_2\,\bar q_3\gamma^\mu(1-\gamma_5)c$    in  $O_\pm^{q_1q_2}$ and $O_\pm^{qq}$
are the four-quark operators,
and the decays of
$O_\pm^{ds}$, $O_\pm^{qq}$ and $O_\pm^{sd}$ are so-called Cabibbo-favored (CF), singly Cabibbo-suppressed (SCS), and
doubly Cabibbo-suppressed (DCS) processes, respectively.

In Eq.~(\ref{O12}), the flavor structure of the four quark operator $(\bar q_1 q_2)_{V-A}(\bar q_3 c)_{V-A}$
%\equiv
%\bar q_1\gamma_\mu(1-\gamma_5)q_2\,\bar q_3\gamma^\mu(1-\gamma_5)c$
can be rewritten as $(\bar q^i q_k \bar q^j)c$ 
with $q_i=(u,d,s)$,  which is the triplet of 3 under the $SU(3)_f$ symmetry, where the Dirac and Lorentz indices are suppressed for simplicity.
%
%Furthermore, 
Since $(\bar q^i q_k \bar q^j)c$ can be decomposed as
the irreducible representations of
$(\bar 3\times 3\times \bar 3)c=(\bar 3+\bar 3'+6+\overline{15})c$, 
one can derive that~\cite{Savage:1989qr}
%,Savage:1991wu}
%
\begin{eqnarray}\label{O_su3}
O_{+(-)}^{ds}\simeq 
&{\cal O}_{\overline{15}(6)}^{ds}&
=\frac{1}{2}(\bar u d\bar s\pm\bar s d\bar u)c\,,\nonumber\\
O_{-(+)}^{qq}\simeq &{\cal O}_{\overline{15}(6)}^{qq}&
=\frac{1}{2}(\bar u d\bar d\pm\bar d d\bar u)c
-\frac{1}{2}(\bar u s\bar s\pm\bar s s\bar u)c\,,\nonumber\\
O_{-(+)}^{sd}\simeq 
&{\cal O}^{sd}_{\overline{15}(6)}&
=\frac{1}{2}(\bar u s\bar d\pm\bar d s\bar u)c\,.
\end{eqnarray}
Consequently,
the effective Hamiltonian in Eq.~(\ref{Heff})
has the expression under the $SU(3)_f$ symmetry, 
given by~\cite{Geng:2017esc,Geng:2018plk,Geng:2017mxn,Geng:2018bow,Geng:2018rse}
\begin{eqnarray}\label{Heff2}
{\cal H}_{eff}&=&\frac{G_F}{\sqrt 2} 
\left[c_-  \frac{\epsilon^{ijl}}{2}H(6)_{lk}+c_+H(\overline{15})_k^{ij}\right]c\,,
\end{eqnarray}
where  $(i,j,k)$ are the flavor indices, while $H(6)$ and $H(\overline{15})$ are symmetric and anti-symmetric flavor tensors of
${\cal O}_{6, \overline{15}}^{(q_1q_2,qq)}$
in Eq.~(\ref{O_su3}), with their non-zero entries
given by~\cite{Savage:1989qr}
%,Savage:1991wu}
\begin{eqnarray}\label{nz_entry}
H(6)_{ij}&=&c_s^2\left(\begin{array}{ccc}
0& 0& 0\\
0&2&2t_c\\
0&2t_c&2t_c^2
\end{array}\right)\,,\nonumber\\
H(\overline{15})^{ij}_{k}&=&c_s^2\left(\begin{array}{ccc} 
\left(\begin{array}{ccc}
0& 0& 0\\
0&0&0\\
0&0&0
\end{array} \right)  \,,
&
\left( \begin{array}{ccc}
0& -t_c& 1\\
-t_c&0&0\\
1&0&0
\end{array} \right)  \,,
&
\left( \begin{array}{ccc}
0& -t_c^2& t_c\\
-t_c^2&0&0\\
t_c& 0&0
\end{array} \right)  
\end{array}
\right)\,,\nonumber\\
\end{eqnarray}
respectively.  
The three lowest-lying charmed baryon states of ${\bf B}_c$
form an anti-triplet of $\bar 3$ to consist of
$(ds-sd)c$, $(us-su)c$ and $(ud-du)c$,
and ${\bf B}_n$ and $M$ belong to octet (8) states of the baryon and meson, which are written as
\begin{eqnarray}\label{BM_SU3}
{\bf B}_{c}&=&(\Xi_c^0,-\Xi_c^+,\Lambda_c^+)\,,\nonumber\\
{\bf B}_n&=&\left(\begin{array}{ccc}
\frac{1}{\sqrt{6}}\Lambda^0+\frac{1}{\sqrt{2}}\Sigma^0 & \Sigma^- & \Xi^-\\
 \Sigma^+ &\frac{1}{\sqrt{6}}\Lambda^0 -\frac{1}{\sqrt{2}}\Sigma^0  & \Xi^0\\
 p & n &-\sqrt{\frac{2}{3}}\Lambda^0
\end{array}\right)\,,\nonumber\\
M&=&\left(\begin{array}{ccc} 
	\frac{1}{\sqrt{2}}\pi^0+\frac{1}{\sqrt{6}}\eta & \pi^- & K^-\\
	\pi^+ &-\frac{1}{\sqrt{2}}\pi^0+\frac{1}{\sqrt{6}}\eta& \bar K^0\\
	K^+ & K^0& -\sqrt{\frac{2}{3}}\eta
\end{array}\right)\,,
\end{eqnarray}
%========================================
respectively.

%Since the Wilson coefficients are scale-dependent,
In the NDR scheme, the Wilson coefficients of $(c_-,c_+)$ are found to be  $(1.78,0.76)$ 
at the scale of $\mu=1$ GeV~\cite{Fajfer:2002gp}, resulting in 
%We assume 
that the amplitudes associated with $H(\overline{15})$
are suppressed due to the value of $(c_-/c_+)^2\simeq 5.5$.
In addition, the nonfactorizable contributions to the decays from  $H(\overline{15})$ are zero due to
the vanishing baryonic transition matrix elements from  ${\bf \overline{15}}$~\cite{Cheng:2018hwl},
while the factorizable are found to be small in most of the modes~\cite{Geng:2018rse}.
For the three-day decay of ${\bf B_c}\to {\bf B_n}MM'$,
in this study we only consider  the s-wave meson-pair  in the final-state configuration, 
 regarded as a off-shell scalar particle inspired. 
 As a result,
the spin dependent decay amplitude can be simply written as
\begin{eqnarray}\label{3b_amp}
{\cal M}({\bf B_c}\to {\bf B_n}MM')= \langle {\bf B_n}MM'|{\cal H}_{eff}|{\bf B_c}\rangle =i\bar{u}_{\bf B_n}(A-B\gamma_5)u_{\bf B_c}
\end{eqnarray}
where $u_{\bf B_{c,n}}$ are Dirac spinors of baryons,
while  $A$ and $B$ 
%are the s-wave(p-wave) amplitudes, 
correspond to the parity conserving and parity violating processes, which can be represented by $SU(3)$ irreducible amplitudes, given by
\begin{eqnarray}\label{su3_amp}
A({\bf B_c}\to {\bf B}_n MM')&=& %\nonumber\\
%&&
a_1(\bar {\bf B}_n)^k_i (M)^m_l(M)^l_mH(6)_{jk}T^{ij}+a_2(\bar {\bf B}_n)^k_i (M)^m_j(M)^l_mH(6)_{kl}T^{ij}\nonumber\\
&&+
a_3(\bar {\bf B}_n)^k_i (M)^m_k(M)^l_mH(6)_{jl}T^{ij}+a_4(\bar {\bf B}_n)^k_i (M)^l_j(M)^m_kH(6)_{lm}T^{ij}\nonumber\\
&&+
a_5(\bar {\bf B}_n)^l_k (M)^m_j(M)^k_mH(6)_{il}T^{ij}+a_6(\bar {\bf B}_n)^l_k (M)^m_j(M)^k_lH(6)_{im}T^{ij}\,, \nonumber \\
B({\bf B_c}\to {\bf B}_n MM')&=& 
A({\bf B_c}\to {\bf B}_n MM')\{a_i\rightarrow b_i\}
\,,
\end{eqnarray}
respectively,
where $T^{ij}=\epsilon^{ijk}({\bf B_c})_{k}$.
We note that since  CP violating effects in the charmed decays are negligible,
the parameters of $a_{i}$ and $b_{i} $ 
can be taken to be  relatively real~\cite{Textbooks}, so that there are totally 12 real  parameters 
in the $SU(3)_f$ irreducible amplitudes  of Eq.~(\ref{su3_amp}).
Here, we have also assumed that the final state interactions can be ignored. 
We remark that
there are some cases in which the  contributions from the s-wave meson-pairs vanish due to the flavor structure and Bose statistics, 
whereas the p-wave ones are dominant, leading to a different set of spin dependent amplitudes,
 which will not be discussed  in this study.
For example, the decay of 
$\Lambda_c^+\to \Lambda\pi^+\pi^0$
with the measured branching ratio around $7.1\%$ 
is predicted mainly from the contribution of the p-wave meson-pair.
% by the $SU(3)_f$ flavor symmetry. 
The explicit expansions of 
$A(\Lambda_c^+\to {\bf B}_n MM')$, $A (\Xi_{c}^{+} \to {\bf B}_n MM')$ and
$A(\Xi_{c}^{0} \to {\bf B}_n MM')$ are presented in Tables~\ref{Tamp1}, \ref{Tamp2} and \ref{Tamp3},
%  in Appendix.
while those of $B({\bf B_c}\to {\bf B}_n MM')$ can be found by replacing  $a_{i}$ in $A({\bf B_c}\to {\bf B}_n MM')$ with $b_{i}$,
 respectively.
\begin{table}[h!]
	\caption{A-amplitudes of $ \Lambda_{c}^{+}  \to {\bf B_n}MM'$.}\label{Tamp1}
	{		\scriptsize
		%
		%\begin{tabular}{ccc}
		\begin{tabular}{|c|c|}
			\hline
			CF mode%Channel  
			& $A$\\
			\hline
			$ \Sigma^{+} \pi^{0} \pi^{0} $ & $ 4a_{1} + 2a_{2} + 2a_{3} + 2a_{4} - 2a_{5} $ \\
			$ \Sigma^{+} \pi^{+} \pi^{-} $ & $ 4a_{1} + 2a_{2} + 2a_{3} - 2a_{5} - 2a_{6} $ \\
			$ \Sigma^{+} K^{0} \bar{K}^{0} $ & $ 4a_{1} + 2a_{2} + 2a_{3} $ \\
			$ \Sigma^{+} K^{+} K^{-} $ & $ 4a_{1} - 2a_{5} $ \\
			$ \Sigma^{+} \eta^{0} \eta^{0} $ & $ 4a_{1} +\frac{ 2a_{2}}{3}+\frac{ 2a_{3}}{3}+\frac{ 2a_{4}}{3}-\frac{ 2a_{5}}{3} $ \\
			$ \Sigma^{0} \pi^{0} \pi^{+} $ & $ -2a_{4} - 2a_{6} $ \\
			$ \Sigma^{0} K^{+} \bar{K}^{0} $ & $ \sqrt{2}a_{2} + \sqrt{2}a_{3} + \sqrt{2}a_{5} $ \\
			$ \Sigma^{-} \pi^{+} \pi^{+} $ & $ -4a_{4} - 4a_{6} $ \\
			$ \Xi^{0} \pi^{0} K^{+} $ & $ -\sqrt{2}a_{5} $ \\
			$ \Xi^{0} \pi^{+} K^{0} $ & $ -2a_{5} - 2a_{6} $ \\
			$ \Xi^{-} \pi^{+} K^{+} $ & $ -2a_{6} $ \\
			$ p \pi^{0} \bar{K}^{0} $ & $ -\sqrt{2}a_{3} - \sqrt{2}a_{4} $ \\
			$ p \pi^{+} K^{-} $ & $ 2a_{3} - 2a_{6} $ \\
			$ p \bar{K}^{0} \eta^{0} $ & $ -\frac{\sqrt{6}a_{3}}{3}+\frac{ \sqrt{6}a_{4}}{3} $ \\
			$ n \pi^{+} \bar{K}^{0} $ & $ -2a_{4} - 2a_{6} $ \\
			$ \Lambda^{0} \pi^{+} \eta^{0} $ & $ -\frac{2a_{2}}{3}+\frac{ 2a_{3}}{3}-\frac{ 2a_{5}}{3}- 2a_{6} $ \\
			$ \Lambda^{0} K^{+} \bar{K}^{0} $ & $ -\frac{\sqrt{6}a_{2}}{3}+\frac{ \sqrt{6}a_{3}}{3}-\frac{ \sqrt{6}a_{5}}{3} $ \\
			\hline
		\end{tabular}
		%&
		\begin{tabular}{|c|c|}
			\hline
			CS mode%Channel  
			& $At_c^{-1}$\\
			
			\hline
			$ \Sigma^{+} \pi^{0} K^{0} $ & $ \sqrt{2}a_{2} + \sqrt{2}a_{3} + 2\sqrt{2}a_{4} $ \\
			$ \Sigma^{+} \pi^{-} K^{+} $ & $ -2a_{2} - 2a_{3} + 2a_{6} $ \\
			$ \Sigma^{+} K^{0} \eta^{0} $ & $ \frac{\sqrt{6}a_{2}}{3}+\frac{ \sqrt{6}a_{3}}{3}-\frac{ 2\sqrt{6}a_{4}}{3} $ \\
			$ \Sigma^{0} \pi^{+} K^{0} $ & $ -\sqrt{2}a_{2} - \sqrt{2}a_{3} - 2\sqrt{2}a_{4} $ \\
			$ \Sigma^{0} K^{+} \eta^{0} $ & $ \frac{\sqrt{3}a_{2}}{3}+\frac{ \sqrt{3}a_{3}}{3}-\frac{ 2\sqrt{3}a_{4}}{3} $ \\
			$ \Sigma^{-} \pi^{+} K^{+} $ & $ 4a_{4} + 2a_{6} $ \\
			$ p \pi^{0} \pi^{0} $ & $ -4a_{1} - 2a_{2} + 2a_{5} $ \\
			$ p \pi^{0} \eta^{0} $ & $ \frac{2\sqrt{3}a_{2}}{3}-\frac{ 2\sqrt{3}a_{4}}{3}+\frac{ 2\sqrt{3}a_{5}}{3} $ \\
			$ p \pi^{+} \pi^{-} $ & $ -4a_{1} - 2a_{2} + 2a_{5} $ \\
			$ p K^{+} K^{-} $ & $ -4a_{1} - 2a_{3} + 2a_{5} + 2a_{6} $ \\
			$ p \eta^{0} \eta^{0} $ & $ -4a_{1} -\frac{ 2a_{2}}{3}-\frac{ 8a_{3}}{3}+\frac{ 4a_{4}}{3}+\frac{ 2a_{5}}{3} $ \\
			$ n \pi^{+} \eta^{0} $ & $ \frac{2\sqrt{6}a_{2}}{3}-\frac{ 2\sqrt{6}a_{4}}{3}+\frac{ 2\sqrt{6}a_{5}}{3} $ \\
			$ n K^{+} \bar{K}^{0} $ & $ 2a_{2} + 2a_{4} + 2a_{5} + 2a_{6} $ \\
			$ \Lambda^{0} \pi^{0} K^{+} $ & $ \frac{\sqrt{3}a_{2}}{3}-\frac{ \sqrt{3}a_{3}}{3}-\frac{ 2\sqrt{3}a_{5}}{3} $ \\
			$ \Lambda^{0} \pi^{+} K^{0} $ & $ \frac{\sqrt{6}a_{2}}{3}-\frac{ \sqrt{6}a_{3}}{3}-\frac{ 2\sqrt{6}a_{5}}{3} $ \\
			$ \Lambda^{0} K^{+} \eta^{0} $ & $ -\frac{a_{2}}{3}+\frac{ a_{3}}{3}+\frac{ 2a_{5}}{3}+ 2a_{6} $ \\
			&\\
			\hline
		\end{tabular}
		%&
		\begin{tabular}{|c|c|}
			\hline
			DCS mode%Channel  
			& $At_c^{-2}$\\
			\hline
			$ \Sigma^{+} K^{0} K^{0} $ & $ 4a_{4} $ \\
			$ \Sigma^{0} K^{0} K^{+} $ & $ 2\sqrt{2}a_{4} $ \\
			$ \Sigma^{-} K^{+} K^{+} $ & $ -4a_{4} $ \\
			$ p \pi^{0} K^{0} $ & $ -\sqrt{2}a_{2} $ \\
			$ p \pi^{-} K^{+} $ & $ 2a_{2} $ \\
			$ p K^{0} \eta^{0} $ & $ -\frac{\sqrt{6}a_{2}}{3}-\frac{ 2\sqrt{6}a_{4}}{3} $ \\
			$ n \pi^{0} K^{+} $ & $ -\sqrt{2}a_{2} $ \\
			$ n \pi^{+} K^{0} $ & $ -2a_{2} $ \\
			$ n K^{+} \eta^{0} $ & $ \frac{\sqrt{6}a_{2}}{3}+\frac{ 2\sqrt{6}a_{4}}{3} $ \\	
			&\\
			&\\
			&\\
			&\\
			&\\
			&\\
			&\\
			&\\
			\hline
		\end{tabular}
		%\end{tabular}
	}
\end{table}
%====================================
%\newpage
%Amp2: ====================================
\begin{table}[h!]
	%	\centering
	\caption{A-amplitudes of $ \Xi_{c}^{+}  \to {\bf B_n}MM'$.}\label{Tamp2}
	{%\tiny
		\scriptsize
		%
		%\begin{tabular}{ccc}
		\begin{tabular}{|c|c|}
			\hline
			CF mode& $A$\\
			\hline
			$ \Sigma^{+} \pi^{0} \bar{K}^{0} $ & $ -\sqrt{2}a_{2} - \sqrt{2}a_{4} $ \\
			$ \Sigma^{+} \pi^{+} K^{-} $ & $ 2a_{2} $ \\
			$ \Sigma^{+} \bar{K}^{0} \eta^{0} $ & $ -\frac{\sqrt{6}a_{2}}{3}+\frac{ \sqrt{6}a_{4}}{3} $ \\
			$ \Sigma^{0} \pi^{+} \bar{K}^{0} $ & $ \sqrt{2}a_{4} $ \\
			$ \Xi^{0} \pi^{0} \pi^{+} $ & $ \sqrt{2}a_{4} $ \\
			$ \Xi^{0} \pi^{+} \eta^{0} $ & $ -\frac{2\sqrt{6}a_{2}}{3}-\frac{ \sqrt{6}a_{4}}{3} $ \\
			$ \Xi^{0} K^{+} \bar{K}^{0} $ & $ -2a_{2} $ \\
			$ \Xi^{-} \pi^{+} \pi^{+} $ & $ -4a_{4} $ \\
			$ p \bar{K}^{0} \bar{K}^{0} $ & $ 4a_{4} $ \\
			$ \Lambda^{0} \pi^{+} \bar{K}^{0} $ & $ \sqrt{6}a_{4} $ \\
			&\\
			&\\
			&\\
			&\\
			&\\
			&\\
			&\\
			&\\
			&\\
			\hline
		\end{tabular}
		%&
		\begin{tabular}{|c|c|}
			\hline
			CS mode& $At_c^{-1}$\\
			\hline
			$ \Sigma^{+} \pi^{0} \pi^{0} $ & $ -4a_{1} - 2a_{3} + 2a_{5} $ \\
			$ \Sigma^{+} \pi^{0} \eta^{0} $ & $ \frac{2\sqrt{3}a_{3}}{3}-\frac{ 2\sqrt{3}a_{4}}{3}+\frac{ 2\sqrt{3}a_{5}}{3} $ \\
			$ \Sigma^{+} \pi^{+} \pi^{-} $ & $ -4a_{1} - 2a_{3} + 2a_{5} + 2a_{6} $ \\
			$ \Sigma^{+} K^{+} K^{-} $ & $ -4a_{1} - 2a_{2} + 2a_{5} $ \\
			$ \Sigma^{+} \eta^{0} \eta^{0} $ & $ -4a_{1} -\frac{ 8a_{2}}{3}-\frac{ 2a_{3}}{3}+\frac{ 4a_{4}}{3}+\frac{ 2a_{5}}{3} $ \\
			$ \Sigma^{0} \pi^{0} \pi^{+} $ & $ 2a_{6} $ \\
			$ \Sigma^{0} \pi^{+} \eta^{0} $ & $ -\frac{2\sqrt{3}a_{3}}{3}+\frac{ 2\sqrt{3}a_{4}}{3}-\frac{ 2\sqrt{3}a_{5}}{3} $ \\
			$ \Sigma^{0} K^{+} \bar{K}^{0} $ & $ -\sqrt{2}a_{3} - \sqrt{2}a_{4} - \sqrt{2}a_{5} $ \\
			$ \Sigma^{-} \pi^{+} \pi^{+} $ & $ 4a_{6} $ \\
			$ \Xi^{0} \pi^{0} K^{+} $ & $ \sqrt{2}a_{2} - \sqrt{2}a_{4} + \sqrt{2}a_{5} $ \\
			$ \Xi^{0} \pi^{+} K^{0} $ & $ 2a_{2} + 2a_{4} + 2a_{5} + 2a_{6} $ \\
			$ \Xi^{0} K^{+} \eta^{0} $ & $ -\frac{\sqrt{6}a_{2}}{3}+\frac{ \sqrt{6}a_{4}}{3}-\frac{ \sqrt{6}a_{5}}{3} $ \\
			$ \Xi^{-} \pi^{+} K^{+} $ & $ 4a_{4} + 2a_{6} $ \\
			$ p \pi^{0} \bar{K}^{0} $ & $ \sqrt{2}a_{2} + \sqrt{2}a_{3} $ \\
			$ p \pi^{+} K^{-} $ & $ -2a_{2} - 2a_{3} + 2a_{6} $ \\
			$ p \bar{K}^{0} \eta^{0} $ & $ \frac{\sqrt{6}a_{2}}{3}+\frac{ \sqrt{6}a_{3}}{3}+\frac{ 4\sqrt{6}a_{4}}{3} $ \\
			$ n \pi^{+} \bar{K}^{0} $ & $ 2a_{6} $ \\
			$ \Lambda^{0} \pi^{+} \eta^{0} $ & $ -\frac{4a_{2}}{3}-\frac{ 2a_{3}}{3}+ 2a_{4} +\frac{ 2a_{5}}{3}+ 2a_{6} $ \\
			$ \Lambda^{0} K^{+} \bar{K}^{0} $ & $ -\frac{2\sqrt{6}a_{2}}{3}-\frac{ \sqrt{6}a_{3}}{3}- \sqrt{6}a_{4} +\frac{ \sqrt{6}a_{5}}{3} $ \\
			\hline
		\end{tabular}
		%&
		\begin{tabular}{|c|c|}
			\hline
			DCS mode& $At_c^{-2}$\\
			\hline
			$ \Sigma^{+} \pi^{0} K^{0} $ & $ -\sqrt{2}a_{3} $ \\
			$ \Sigma^{+} \pi^{-} K^{+} $ & $ 2a_{3} - 2a_{6} $ \\
			$ \Sigma^{+} K^{0} \eta^{0} $ & $ -\frac{\sqrt{6}a_{3}}{3}-\frac{ 2\sqrt{6}a_{4}}{3} $ \\
			$ \Sigma^{0} \pi^{0} K^{+} $ & $ a_{3} - 2a_{6} $ \\
			$ \Sigma^{0} \pi^{+} K^{0} $ & $ \sqrt{2}a_{3} $ \\
			$ \Sigma^{0} K^{+} \eta^{0} $ & $ -\frac{\sqrt{3}a_{3}}{3}-\frac{ 2\sqrt{3}a_{4}}{3} $ \\
			$ \Sigma^{-} \pi^{+} K^{+} $ & $ -2a_{6} $ \\
			$ \Xi^{0} K^{0} K^{+} $ & $ -2a_{4} - 2a_{6} $ \\
			$ \Xi^{-} K^{+} K^{+} $ & $ -4a_{4} - 4a_{6} $ \\
			$ p \pi^{0} \pi^{0} $ & $ 4a_{1} - 2a_{5} $ \\
			$ p \pi^{0} \eta^{0} $ & $ -\frac{2\sqrt{3}a_{5}}{3} $ \\
			$ p \pi^{+} \pi^{-} $ & $ 4a_{1} - 2a_{5} $ \\
			$ p K^{0} \bar{K}^{0} $ & $ 4a_{1} + 2a_{2} + 2a_{3} $ \\
			$ p K^{+} K^{-} $ & $ 4a_{1} + 2a_{2} + 2a_{3} - 2a_{5} - 2a_{6} $ \\
			$ p \eta^{0} \eta^{0} $ & $ 4a_{1} +\frac{ 8a_{2}}{3}+\frac{ 8a_{3}}{3}+\frac{ 8a_{4}}{3}-\frac{ 2a_{5}}{3} $ \\
			$ n \pi^{+} \eta^{0} $ & $ -\frac{2\sqrt{6}a_{5}}{3} $ \\
			$ n K^{+} \bar{K}^{0} $ & $ -2a_{5} - 2a_{6} $ \\
			$ \Lambda^{0} \pi^{0} K^{+} $ & $ \frac{2\sqrt{3}a_{2}}{3}+\frac{ \sqrt{3}a_{3}}{3}+\frac{ 2\sqrt{3}a_{5}}{3} $ \\
			$ \Lambda^{0} \pi^{+} K^{0} $ & $ \frac{2\sqrt{6}a_{2}}{3}+\frac{ \sqrt{6}a_{3}}{3}+\frac{ 2\sqrt{6}a_{5}}{3} $ \\
			
			\hline
			%\end{tabular}
		\end{tabular}
	}
\end{table}
%====================

%Amp3: ====================
\begin{table}[h!]
	%\centering
	\caption{A-amplitudes of $ \Xi_{c}^{0}  \to {\bf B_n}MM'$.}\label{Tamp3}
	{%\tiny
		\scriptsize
		%
		%\begin{tabular}{cc}
		\begin{tabular}{|c|c|}
			\hline
			CF mode& $A$\\
			\hline
			$ \Sigma^{+} \pi^{0} K^{-} $ & $ \sqrt{2}a_{5} $ \\
			$ \Sigma^{+} \pi^{-} \bar{K}^{0} $ & $ 2a_{5} + 2a_{6} $ \\
			$ \Sigma^{+} K^{-} \eta^{0} $ & $ -\frac{\sqrt{6}a_{5}}{3} $ \\
			$ \Sigma^{0} \pi^{0} \bar{K}^{0} $ & $ a_{2} + a_{4} + a_{5} + 2a_{6} $ \\
			$ \Sigma^{0} \pi^{+} K^{-} $ & $ -\sqrt{2}a_{2} - \sqrt{2}a_{5} $ \\
			$ \Sigma^{0} \bar{K}^{0} \eta^{0} $ & $ \frac{\sqrt{3}a_{2}}{3}-\frac{ \sqrt{3}a_{4}}{3}+\frac{ \sqrt{3}a_{5}}{3} $ \\
			$ \Sigma^{-} \pi^{+} \bar{K}^{0} $ & $ 2a_{4} + 2a_{6} $ \\
			$ \Xi^{0} \pi^{0} \eta^{0} $ & $ \frac{2\sqrt{3}a_{2}}{3}+\frac{ 2\sqrt{3}a_{3}}{3}+\frac{ 2\sqrt{3}a_{4}}{3} $ \\
			$ \Xi^{0} \pi^{+} \pi^{-} $ & $ -4a_{1} - 2a_{2} - 2a_{3} $ \\
			$ \Xi^{0} K^{0} \bar{K}^{0} $ & 
			$-2(2a_{1} +a_{2}+a_{3}$\\
			&$-a_{5}-a_{6})$ \\
			$ \Xi^{0} K^{+} K^{-} $ & $ -4a_{1} + 2a_{5} $ \\
			$ \Xi^{0} \eta^{0} \eta^{0} $ & 
			$-2(2a_{1}+\frac{a_{2}}{3}+\frac{a_{3}}{3}$\\
			&$+\frac{a_{4}}{3}-\frac{4a_{5}}{3})$ \\
			$ \Xi^{-} \pi^{0} \pi^{+} $ & $ \sqrt{2}a_{4} $ \\
			$ \Xi^{-} \pi^{+} \eta^{0} $ & $ -\frac{2\sqrt{6}a_{3}}{3}-\frac{ \sqrt{6}a_{4}}{3} $ \\
			$ \Xi^{-} K^{+} \bar{K}^{0} $ & $ -2a_{3} + 2a_{6} $ \\
			$ p K^{-} \bar{K}^{0} $ & $ 2a_{6} $ \\
			$ n \bar{K}^{0} \bar{K}^{0} $ & $ 4a_{4} + 4a_{6} $ \\
			$ \Lambda^{0} \pi^{0} \bar{K}^{0} $ & 
			$-\sqrt{3}(\frac{a_{2}}{3}+\frac{ 2a_{3}}{3}+a_{4}+\frac{a_{5}}{3})$ \\
			$ \Lambda^{0} \pi^{+} K^{-} $ & $ \frac{\sqrt{6}a_{2}}{3}+\frac{ 2\sqrt{6}a_{3}}{3}+\frac{ \sqrt{6}a_{5}}{3} $ \\
			&\\
			&\\
			&\\
			&\\
			&\\
			&\\
			&\\
			&\\
			\hline
		\end{tabular}
		\begin{tabular}{|c|c|}
			\hline
			CS mode& $At_c^{-1}$ \\
			\hline
			$ \Sigma^{+} \pi^{0} \pi^{-} $ & $ -\sqrt{2}a_{6} $ \\
			$ \Sigma^{+} \pi^{-} \eta^{0} $ & $ \frac{2\sqrt{6}a_{5}}{3}+ \sqrt{6}a_{6} $ \\
			$ \Sigma^{+} K^{0} K^{-} $ & $ 2a_{5} $ \\
			$ \Sigma^{0} \pi^{0} \pi^{0} $ & $ 2\sqrt{2}a_{1} + \sqrt{2}a_{3} - \sqrt{2}a_{5} - 2\sqrt{2}a_{6} $ \\
			$ \Sigma^{0} \pi^{0} \eta^{0} $ & $ -\frac{\sqrt{6}a_{3}}{3}+\frac{ \sqrt{6}a_{4}}{3}+\frac{ \sqrt{6}a_{5}}{3}+ \sqrt{6}a_{6} $ \\
			$ \Sigma^{0} \pi^{+} \pi^{-} $ & $ 2\sqrt{2}a_{1} + \sqrt{2}a_{3} - \sqrt{2}a_{5} $ \\
			$ \Sigma^{0} K^{0} \bar{K}^{0} $ & 
			$ \sqrt{2}(2a_{1} +a_{2} +a_{3} +a_{4} -a_{5})$ \\
			$ \Sigma^{0} K^{+} K^{-} $ & $ 2\sqrt{2}a_{1} + \sqrt{2}a_{2} $ \\
			$ \Sigma^{0} \eta^{0} \eta^{0} $ & 
			$\sqrt{2}(2a_{1} +\frac{ 4a_{2}}{3}+\frac{a_{3}}{3}-\frac{ 2a_{4}}{3}-\frac{a_{5}}{3})$ \\
			$ \Sigma^{-} \pi^{0} \pi^{+} $ & $ -\sqrt{2}a_{6} $ \\
			$ \Sigma^{-} \pi^{+} \eta^{0} $ & $ -\frac{2\sqrt{6}a_{3}}{3}+\frac{ 2\sqrt{6}a_{4}}{3}+ \sqrt{6}a_{6} $ \\
			$ \Sigma^{-} K^{+} \bar{K}^{0} $ & $ -2a_{3} - 2a_{4} $ \\
			$ \Xi^{0} \pi^{-} K^{+} $ & $ 2a_{2} + 2a_{3} + 2a_{5} $ \\
			$ \Xi^{0} K^{0} \eta^{0} $ & 
			$\sqrt{6}(-\frac{a_{2}}{3}-\frac{a_{3}}{3}
			+\frac{ 2a_{4}}{3}-\frac{a_{5}}{3}+a_{6})$ \\
			$ \Xi^{-} \pi^{0} K^{+} $ & $ \sqrt{2}a_{3} - \sqrt{2}a_{4} - \sqrt{2}a_{6} $ \\
			$ \Xi^{-} \pi^{+} K^{0} $ & $ 2a_{3} + 2a_{4} $ \\
			$ p \pi^{0} K^{-} $ & $ -\sqrt{2}a_{5} - \sqrt{2}a_{6} $ \\
			$ p \pi^{-} \bar{K}^{0} $ & $ -2a_{5} $ \\
			$ p K^{-} \eta^{0} $ & $ \frac{\sqrt{6}a_{5}}{3}+ \sqrt{6}a_{6} $ \\
			$ n \pi^{0} \bar{K}^{0} $ & $ \sqrt{2}a_{2} + \sqrt{2}a_{3} + \sqrt{2}a_{5} - \sqrt{2}a_{6} $ \\
			$ n \pi^{+} K^{-} $ & $ -2a_{2} - 2a_{3} - 2a_{5} $ \\
			$ n \bar{K}^{0} \eta^{0} $ & 
			$\sqrt{6}(\frac{a_{2}}{3}+\frac{a_{3}}{3}+\frac{ 4a_{4}}{3}+\frac{a_{5}}{3}+a_{6})$ \\
			$ \Lambda^{0} \pi^{0} \pi^{0} $ & 
			$\sqrt{6}(-2a_{1} -\frac{ 2a_{2}}{3}-\frac{a_{3}}{3}+\frac{a_{5}}{3})$ \\
			$ \Lambda^{0} \pi^{0} \eta^{0} $ & 
			$\sqrt{2}(\frac{2a_{2}}{3}+\frac{a_{3}}{3}-a_{4} -\frac{a_{5}}{3}-a_{6})$ \\
			$ \Lambda^{0} \pi^{+} \pi^{-} $ & 
			$\sqrt{6}(-2a_{1} -\frac{ 2a_{2}}{3}-\frac{a_{3}}{3}+\frac{a_{5}}{3})$ \\
			$ \Lambda^{0} K^{0} \bar{K}^{0} $ & 
			$\sqrt{6}(-2a_{1} - a_{2} - a_{3} - a_{4} +a_{5})$ \\
			$ \Lambda^{0} K^{+} K^{-} $ & 
			$\sqrt{6}(-2a_{1} -\frac{a_{2}}{3}-\frac{ 2a_{3}}{3}+\frac{ 2a_{5}}{3})$ \\
			$ \Lambda^{0} \eta^{0} \eta^{0} $ & 
			$\sqrt{6}(-2a_{1} -\frac{ 2a_{2}}{3}-a_{3}+\frac{ 2a_{4}}{3}$\\
			&$+a_{5} + 2a_{6})$ \\
			\hline
		\end{tabular}
		\begin{tabular}{|c|c|}
			\hline
			DCS mode& $At_c^{-2}$\\
			\hline
			$ \Sigma^{+} \pi^{-} K^{0} $ & $ -2a_{6} $ \\
			$ \Sigma^{0} \pi^{0} K^{0} $ & $ a_{3} - 2a_{6} $ \\
			$ \Sigma^{0} \pi^{-} K^{+} $ & $ -\sqrt{2}a_{3} $ \\
			$ \Sigma^{0} K^{0} \eta^{0} $ & $ \frac{\sqrt{3}a_{3}}{3}+\frac{ 2\sqrt{3}a_{4}}{3} $ \\
			$ \Sigma^{-} \pi^{0} K^{+} $ & $ \sqrt{2}a_{3} $ \\
			$ \Sigma^{-} \pi^{+} K^{0} $ & $ 2a_{3} - 2a_{6} $ \\
			$ \Sigma^{-} K^{+} \eta^{0} $ & $ -\frac{\sqrt{6}a_{3}}{3}-\frac{ 2\sqrt{6}a_{4}}{3} $ \\
			$ \Xi^{0} K^{0} K^{0} $ & $ -4a_{4} - 4a_{6} $ \\
			$ \Xi^{-} K^{0} K^{+} $ & $ -2a_{4} - 2a_{6} $ \\
			$ p \pi^{-} \eta^{0} $ & $ -\frac{2\sqrt{6}a_{5}}{3} $ \\
			$ p K^{0} K^{-} $ & $ -2a_{5} - 2a_{6} $ \\
			$ n \pi^{0} \pi^{0} $ & $ 4a_{1} - 2a_{5} $ \\
			$ n \pi^{0} \eta^{0} $ & $ \frac{2\sqrt{3}a_{5}}{3} $ \\
			$ n \pi^{+} \pi^{-} $ & $ 4a_{1} - 2a_{5} $ \\
			$ n K^{0} \bar{K}^{0} $ & 
			$2(2a_{1} + a_{2} + a_{3}$\\
			&$ - a_{5} - a_{6})$ \\
			$ n K^{+} K^{-} $ & $ 4a_{1} + 2a_{2} + 2a_{3} $ \\
			$ n \eta^{0} \eta^{0} $ & 
			$ 4a_{1} +\frac{ 8a_{2}}{3}+\frac{ 8a_{3}}{3}$\\
			&$+\frac{ 8a_{4}}{3}-\frac{ 2a_{5}}{3} $ \\
			$ \Lambda^{0} \pi^{0} K^{0} $ & $-\sqrt{3}(\frac{2a_{2}}{3}+\frac{a_{3}}{3}+\frac{ 2a_{5}}{3})$ \\
			$ \Lambda^{0} \pi^{-} K^{+} $ & 
			$\sqrt{6}(\frac{2a_{2}}{3}+\frac{a_{3}}{3}+\frac{ 2a_{5}}{3})$ \\
			&\\
			&\\
			&\\
			&\\
			&\\
			&\\
			&\\
			&\\
			\hline
		\end{tabular}
	}
\end{table}

The differential decay width with an unpolarized  $\bf B_c$ and up-down asymmetry $\alpha$ 
in ${\bf B_c}\to {\bf B_n}MM'$
are given by
\begin{eqnarray}
d\Gamma&=&\frac{1}{(2\pi)^3}\frac{|S|^2+|P|^2+2Re(S^{*}P)\vec{s}_{\bf Bn} \cdot \vec{p}_{\bf B_n}  }{64m^3_{\bf B_c}}
dm_{12}^2 dm_{23}^2\,, 
%\nonumber \\
\end{eqnarray}
and
\begin{eqnarray}
\alpha&=&\frac{d\Gamma(\vec{s}_{\bf Bn} \cdot \vec{p}_{\bf B_n}=1)-d\Gamma(\vec{s}_{\bf Bn} \cdot \vec{p}_{\bf B_n}=-1)}{d\Gamma(\vec{s}_{\bf Bn} \cdot \vec{p}_{\bf B_n}=1)+d\Gamma(\vec{s}_{\bf Bn} \cdot \vec{p}_{\bf B_n}=-1)}=\frac{2Re(S^*P)}{|S|^2+|P|^2}\,,
% \nonumber\\
\end{eqnarray}
respectively,
where
\begin{eqnarray}
S&=&A \,,\quad P=\sqrt{\frac{(m_{\bf B_c}-m_{\bf B_n})^2-m_{23}^2}{(m_{\bf B_c}+m_{\bf B_n})^2-m_{23}^2}}B=\kappa(m_{23}^2)B\,,
\end{eqnarray}
with $m_{12}=p_{\bf B_n}+p_{M'}$, $m_{23}=p_{M}+p_{M'}$
and $\kappa^2(m_{23}^2)=((m_{\bf B_c}-m_{\bf B_n})^2-m_{23}^2)/((m_{\bf B_c}+m_{\bf B_n})^2-m_{23}^2)$.
In general, $A$ and $B$  in Eq.~(\ref{su3_amp}) depend on  $m_{12}^2$ and $m_{23}^2$. 
However, we can assume them to be constant  when  the  non-resonant contributions are excluded.
Consequently, the decay width $\Gamma$ and averaged up-down asymmetry $\langle \alpha \rangle$ can be derived as follows: 
\begin{eqnarray}
\Gamma&=&\int_{m_{12}^2}\int_{m_{23}^2} d \Gamma(\vec{s}_{\bf Bn} \cdot \vec{p}_{\bf B_n}=1)+ d \Gamma(\vec{s}_{\bf Bn} \cdot \vec{p}_{\bf B_n}=-1)\nonumber\\
&=&\int_{m_{12}^2}\int_{m_{23}^2}\frac{1}{(2\pi)^3}\frac{|S|^2+|P|^2}{32m^3_{\bf B_c}}
dm_{12}^2 dm_{23}^2\,, 
%\nonumber \\
\end{eqnarray}
and
\begin{eqnarray}
\langle \alpha \rangle &\equiv& \frac{\int_{m_{12}^2}\int_{m_{23}^2}d\Gamma(\vec{s}_{\bf Bn} \cdot \vec{p}_{\bf B_n}=1)-d\Gamma(\vec{s}_{\bf Bn} \cdot \vec{p}_{\bf B_n}=-1)}{\int_{m_{12}^2}\int_{m_{23}^2}d\Gamma(\vec{s}_{\bf Bn} \cdot \vec{p}_{\bf B_n}=1)+d\Gamma(\vec{s}_{\bf Bn} \cdot \vec{p}_{\bf B_n}=-1)}\,.
\end{eqnarray}

\section{Numerical results }
In the numerical analysis, we perform 
the minimum $\chi^2$ fit to obtain the values of $a_i$ and $b_i$ in in Eq.~(\ref{su3_amp})  under $SU(3)_f$
for the ${\bf B_c}\to {\bf B}_n MM'$ decays.
The  $\chi^2$ fit  is given by
\begin{eqnarray}
\chi^2=\sum_{i} \bigg(\frac{{\cal B}^i_{SU(3)}-{\cal B}^i_{data}}{\sigma_{data}^i}\bigg)^2\,,
\end{eqnarray}
where  ${\cal B}_{SU(3)}^i$  represents the $i$-th decay branching ratio from
the $SU(3)_f$ irreducible amplitude,  ${\cal B}_{data}^i$ stands for the $i$-th experimental data,
and $\sigma_{data}^i$ corresponds to the $i$-th experimental error, while
$i=1,2,\cdots, 16$  for the 16 measured modes  in Table~\ref{exp}.  
Using $\sin\theta_c=0.2248$~\cite{pdg}, one gets
that $t_c=0.2307$ in Eq.~(\ref{nz_entry}).

We now discuss our  data input sections in Table~\ref{exp}.
First of all, we exclude  the resonant contributions from all the data in the table.
In particular,  we use the non-resonant data of $\Lambda_c^+\to pK^-\pi^+$ 
from the PDG~\cite{pdg}. 
In addition, we pick up the data for $\Lambda_c^+\to p K^- K^+$ and $\Xi_c^0\to \Lambda^{0} K^- K^+$  
without the contributions of the resonant process of $\phi \to K^+K^-$. 
For the other $\Lambda_c^+$ decays, their resonant contributions can be taken to be small
so that they are insensitive to our fitting results, such as 
${\cal B}(\Lambda_c^+\to 
\Sigma^+(\rho^0\to) \pi^+\pi^-)<1.7\%$~\cite{pdg}.
For this reason, we choose the total branching ratios as our data points.
The value of ${\cal B}(\Xi_c^{+}\to \Xi^-\pi^+\pi^+)$ is extracted from the ratio $\frac{{\cal B}(\Xi_{c}^+ \to \Xi^0 e^+ \mu_e )}{{\cal B}(\Xi_c^{+}\to \Xi^-\pi^+\pi^+)}=2.3^{+0.7}_{-0.8}$ in the PDG~\cite{pdg} with the theoretical prediction of
${\cal B}(\Xi_{c}^+ \to \Xi^0 e^+ \mu_e)=(10.8\pm0.9)\%$ by $SU(3)_f$ and heavy quark symmetry~\cite{Geng:2019bfz}.
For $\Xi_c^0$ decay  processes, the data of ${\cal B}(\Xi_{c}^0 \to \Xi^-\pi^+)=(1.8\pm0.5)\%$ measured by BELLE~\cite{Li:2018qak} 
and the ratios of $\frac{{\cal B}(\Xi_{c}^0 \to \Lambda^0 K^- \pi^+)}{{\cal B}(\Xi_c^{0}\to \Xi^-\pi^+)}=1.07\pm0.14$  and $\frac{{\cal B}(\Xi_{c}^0 \to \Lambda^0 K^+ K^- )}{{\cal B}(\Xi_c^{0}\to \Xi^-\pi^+)}=0.029\pm0.007$ in PDG~\cite{pdg} are used to extract the absolute branching ratios
of ${\cal B}(\Xi_{c}^0 \to \Lambda^0 K^- \pi^+)$ and ${\cal B}(\Xi_{c}^0 \to \Lambda^0 K^+ K^- )$.

There are 12 parameters to be extracted 
with 16 data inputs as shown in Table~\ref{exp}.
In  Table~\ref{su3_fit}, we present the fitting values of $a_i$  and $b_i$. 
The correlation coefficients of $i$-th and $j$-th irreducible amplitudes are given by

{%\footnotesize
\scriptsize
\begin{eqnarray}\label{cor}
R=\left(\begin{array}{cccccccccccc}
1&-0.58&-0.44&0.43&0.96&-0.32&-0.01&0.47&-0.33&0.55&0.73&-0.56\\
-0.58&1&0.55&-0.64&-0.49&0.29&0.65&-0.82&0.21&-0.72&-0.79&0.63\\
-0.44&0.55&1&-0.68&-0.36&0.66&0.44&-0.54&0.37&-0.63&-0.45&0.61\\
0.43&-0.64&-0.68&1&0.46&-0.61&-0.60&0.61&-0.33&0.78&0.58&-0.65\\
0.96&-0.49&-0.36&0.46&1&-0.27&0.00&0.41&-0.23&0.55&0.74&-0.52\\
-0.32&0.29&0.66&-0.61&-0.27&1&0.28&-0.34&0.50&-0.46&-0.23&0.54\\
-0.01&0.65&0.44&-0.60&0.00&0.28&1&-0.79&0.38&-0.76&-0.44&0.69\\
0.47&-0.82&-0.54&0.61&0.41&-0.34&-0.79&1&-0.34&0.74&0.82&-0.69\\
-0.33&0.21&0.37&-0.33&-0.23&0.50&0.38&-0.34&1&-0.63&0.01&0.78\\
0.55&-0.72&-0.63&0.78&0.55&-0.46&-0.76&0.74&-0.63&1&0.58&-0.94\\
0.73&-0.79&-0.45&0.58&0.74&-0.23&-0.44&0.82&0.01&0.58&1&-0.48\\
-0.56&0.63&0.61&-0.65&-0.52&0.54&0.69&-0.69&0.78&-0.94&-0.48&1\\
\end{array} \right)\,.
% \nonumber
\end{eqnarray}
}
In our fit, we find that  $\chi^2/d.o.f=9.6/4=2.4$ with $d.o.f$ representing  degree of freedom.
As seen from Table~\ref{exp}, 
 the decay branching ratios are reproduced, which agree well with the data in Refs.~\cite{pdg,Berger:2018pli,Li:2018qak,Aaij:2017rin,BESIII} accordingly.
%%======================================
\begin{table}
\caption{The data inputs from  Refs.~\cite{pdg,Berger:2018pli,Li:2018qak,Aaij:2017rin,BESIII} and reproductions for ${\cal B}(\Lambda_c^+\to {\bf B_n}MM)$.}
\label{exp}
\begin{tabular}{|c|c|c|}
\hline
&data&our results\\
\hline
$10^2{\cal B}(\Lambda_c^+\to pK^-\pi^+)$
&$3.4\pm0.4$&$3.4\pm0.5$\\
%$10^2{\cal B}(\Lambda_c^+\to p\bar K^0\pi^0)$
%&$4.0\pm0.3$&$**\pm**$\\
%&$1.99\pm0.13$\\
$10^3{\cal B}(\Lambda_c^+\to \Lambda^0 K^+\bar K^0)$
&$5.6\pm1.1$&$5.8\pm1.0$\\
$10^2{\cal B}(\Lambda_c^+\to \Lambda^0 \pi^+\eta)$
&$1.8\pm0.3$&$1.7\pm0.3$\\
$10^2{\cal B}(\Lambda_c^+\to \Sigma^+\pi^+\pi^-)$
&$4.4\pm0.3$&$4.5\pm0.3$\\
$10^2{\cal B}(\Lambda_c^+\to \Sigma^-\pi^+\pi^+)$
&$1.9\pm0.2$&$1.9\pm0.3$\\
$10^2{\cal B}(\Lambda_c^+\to \Sigma^0\pi^+\pi^0)$
&$2.2\pm0.8$&$1.0\pm0.1$\\

$10^3{\cal B}(\Lambda_c^+\to \Sigma^+ K^+ \pi^-)$
&$2.1\pm0.6$&$2.5\pm0.3$\\
$10^3{\cal B}(\Lambda_c^+\to \Xi^-K^+\pi^+)$
&$6.2\pm0.6$&$6.1\pm0.8$\\ 
%&$4.0\pm 1.0$&$6.2\pm0.5$\\ %\text{no $\Xi(1530)^0$})\\
$10^3{\cal B}(\Lambda_c^+\to p\pi^-\pi^+)$
&$4.2\pm 0.4$&$4.7\pm0.4$\\
$10^4{\cal B}(\Lambda_c^+\to pK^-K^+)$
&$5.2\pm1.2$&$5.0\pm1.2$\\
\hline
\end{tabular}
\begin{tabular}{|c|c|c|}
\hline
&data&our results\\
\hline
$10^2{\cal B}(\Lambda_c^+\to p\bar K^0\eta)$
&$1.6\pm0.4$&$0.7\pm0.1$\\
$10^2{\cal B}(\Lambda_c^+\to \Sigma^+\pi^0\pi^0)$
&$1.3\pm0.1$&$1.3\pm0.2$\\%~\cite{Berger:2018pli}\\ %&$1.36\pm0.14$\\
$10^4{\cal B}(\Lambda_c^+\to pK^+\pi^-)$
&$1.0\pm 0.1$&$1.0\pm0.1$\\%~\cite{Aaij:2017rin}\\
$10^2{\cal B}(\Xi_c^+\to \Xi^-\pi^+\pi^+)$
&$4.7\pm 1.7$&$5.4\pm1.3$\\
$10^{2}{\cal B}(\Xi_c^0\to \Lambda^0 K^-\pi^+)$
&$1.9\pm0.6$&$2.2\pm0.6$\\
$10^{4}{\cal B}(\Xi_c^0\to \Lambda^0 K^-K^+)$
&$5.2\pm1.9$&$6.2\pm1.2$\\
&&\\
&&\\
&&\\
&&\\
\hline
\end{tabular}
\end{table}
%==================================================
\begin{table}[h]
	\caption{Fitting results for $a_i$ and $b_{i}$ in  unit of GeV$^2$. }\label{su3_fit}
\begin{tabular}{|c|c||c|c|}
	\hline
	$a_i$& result &	$b_i$&result  \\
	\hline
	$a_1$&$9.2\pm 0.7$&$b_1$&$18.3\pm0.9$ \\
		\hline
	$a_2$&$-3.7\pm 0.5$&$b_2$&$-9.8\pm2.4$\\
		\hline
	$a_3$&$-7.3\pm 0.4$&$b_3$&$4.4\pm2.1$ \\
		\hline
	$a_4$&$2.3\pm 0.4$&$b_4$&$-5.4\pm2.9$\\
		\hline
	$a_5$&$11.5\pm 1.3$&$b_5$&$38.8\pm2.2$\\
		\hline
	$a_6$&$-3.7\pm 0.2$&$b_6$&$12.7\pm2.3$ \\
	\hline
\end{tabular}
\end{table}
%from the PDG~\cite{pdg}, except for ${\cal B}(\Lambda_c^+\to 
%\Sigma^+\pi^0\pi^0,pK^+\pi^-)$~\cite{Berger:2018pli,Aaij:2017rin}.\label{exp}
%\section{Discussions and Conclusions}
In Tables~\ref{pre_Lc}, \ref{pre_Xicp} and \ref{pre_Xic0}, 
we show our numerical results for 
the decay  branching ratios in $\Lambda_{c}^{+} \to {\bf B_n}MM'$, $\Xi_{c}^{+} \to {\bf B_n}MM'$ and
$\Xi_{c}^{0} \to {\bf B_n}MM'$, respectively.
\begin{table}[h]
\caption{Numerical results for 
 ${\cal B}(\Lambda_{c}^{+} \to {\bf B_n}MM'$).}\label{pre_Lc}
 {\scriptsize
\begin{tabular}{|c|c|}
\hline
CF mode  & $10^{3}{\cal B}$\\
\hline
%$10^{ 2 }{\cal B}_{ \Sigma^{+} \pi^{0} \pi^{0} }$ & $     1.2635 \pm     0.1588 $ \\
$\Sigma^{+} \pi^{0} \eta^{0} $ & $     6.6 \pm     3.4 $ \\
%$10^{ 2 }{\cal B}_{ \Sigma^{+} \pi^{+} \pi^{-} }$ & $     4.4501 \pm     0.3232 $ \\
$\Sigma^{+} K^{0} \bar{K}^{0} $ & $     2.9 \pm     0.7 $ \\
$\Sigma^{+} K^{+} K^{-} $ & $     2.5 \pm     0.3 $ \\
$ \Sigma^{+} \eta^{0} \eta^{0} $ & $    ( 3.2 \pm     0.4)\times10^{-4} $ \\
%$10^{ 2 }{\cal B}_{ \Sigma^{0} \pi^{0} \pi^{+} }$ & $     0.9656 \pm     0.1343 $ \\
$\Sigma^{0} \pi^{+} \eta^{0} $ & $     6.3 \pm     3.2 $ \\
$ \Sigma^{0} K^{+} \bar{K}^{0} $ & $    0.26 \pm     0.09 $ \\
%$10^{ 2 }{\cal B}_{ \Sigma^{-} \pi^{+} \pi^{+} }$ & $     1.8715 \pm     0.2635 $ \\
$ \Xi^{0} \pi^{0} K^{+} $ & $     32 \pm     6 $ \\
$ \Xi^{0} \pi^{+} K^{0} $ & $     44 \pm     8 $ \\
%$10^{ 2 }{\cal B}_{ \Xi^{-} \pi^{+} K^{+} }$ & $     0.6163 \pm     0.0788 $ \\
$ p \pi^{0} \bar{K}^{0} $ & $     23 \pm     4 $ \\
%$10^{ 2 }{\cal B}_{ p \pi^{+} K^{-} }$ & $     3.3896 \pm     0.4896 $ \\
%$10^{ 2 }{\cal B}_{ p \bar{K}^{0} \eta^{0} }$ & $     0.6740 \pm     0.1078 $ \\
$ n \pi^{+} \bar{K}^{0} $ & $     11 \pm     1 $ \\
%$10^{ 2 }{\cal B}_{ \Lambda^{0} \pi^{+} \eta^{0} }$ & $     1.7029 \pm     0.2704 $ \\
%$10^{ 2 }{\cal B}_{ \Lambda^{0} K^{+} \bar{K}^{0} }$ & $     0.5808 \pm     0.1008 $ \\

&\\
&\\
&\\
&\\
\hline
\end{tabular}
\begin{tabular}{|c|c|}
\hline
CS mode  & $10^{4}{\cal B}$\\
\hline
$ \Sigma^{+} \pi^{0} K^{0} $ & $     9.9 \pm     2.8$ \\
%$10^{ 4 }{\cal B}_{ \Sigma^{+} \pi^{-} K^{+} }$ & $    25.1836 \pm     2.5785 $ \\
$\Sigma^{+} K^{0} \eta^{0} $ & $     0.26 \pm     0.06 $ \\
$ \Sigma^{0} \pi^{0} K^{+} $ & $     7.8 \pm     2.3 $ \\
$ \Sigma^{0} \pi^{+} K^{0} $ & $     9.6 \pm     2.7 $ \\
$ \Sigma^{0} K^{+} \eta^{0} $ & $     0.13 \pm     0.03 $ \\
$ p \pi^{0} \pi^{0} $ & $    24 \pm     2 $ \\
$ p \pi^{0} \eta^{0} $ & $    34 \pm     7 $ \\
%$10^{ 4 }{\cal B}_{ p \pi^{+} \pi^{-} }$ & $    47.0477 \pm     3.5243 $ \\
$ p K^{0} \bar{K}^{0} $ & $    37 \pm     8 $ \\
%$10^{ 4 }{\cal B}_{ p K^{+} K^{-} }$ & $     5.0292 \pm     1.2059 $ \\
$ p \eta^{0} \eta^{0} $ & $     2.8 \pm     1.2$ \\
$ n \pi^{+} \eta^{0} $ & $    67 \pm    13 $ \\
$n K^{+} \bar{K}^{0} $ & $    31 \pm     9 $ \\
$ \Lambda^{0} \pi^{0} K^{+}$ & $    35 \pm     6 $ \\
$\Lambda^{0} \pi^{+} K^{0} $ & $    67 \pm    11 $ \\
$ \Lambda^{0} K^{+} \eta^{0} $ & $     0.45 \pm     0.10 $ \\
\hline
\end{tabular}
\begin{tabular}{|c|c|}
\hline
DCS mode&$10^{6}{\cal B}$\\
\hline
$ \Sigma^{+} K^{0} K^{0} $ & $     1.3 \pm     0.5 $ \\
$ \Sigma^{0} K^{0} K^{+}$ & $     1.3 \pm     0.5 $ \\
$ \Sigma^{-} K^{+} K^{+} $ & $     1.3 \pm     0.5 $ \\
$ p \pi^{0} K^{0} $ & $    50 \pm     6 $ \\
%$ p \pi^{-} K^{+} }$ & $   100.5518 \pm    11.3166 $ \\
$ p K^{0} \eta^{0} $ & $     3.3 \pm     2.7 $ \\
$n \pi^{0} K^{+} $ & $    51 \pm     6 $ \\
$ n \pi^{+} K^{0} $ & $    99 \pm    11 $ \\
$n K^{+} \eta^{0} $ & $     3.4 \pm     2.7 $ \\
&   \\
&   \\
&   \\
&   \\
&   \\	
&   \\														
\hline
\end{tabular}
}
\end{table}

\begin{table}
\caption{
Numerical results for 
 ${\cal B}(\Xi_{c}^{+} \to {\bf B_n}MM'$).}\label{pre_Xicp}
%\begin{tabular}{ccc}	
 {\scriptsize
\begin{tabular}{|c|c|}
\hline
CF mode& $10^{2}{\cal B}$\\
\hline
$ \Sigma^{+} \pi^{0} \bar{K}^{0} $ & $     2.0 \pm     1.3 $ \\
$\Sigma^{+} \pi^{+} K^{-} $ & $     5.4 \pm     0.5 $ \\
$ \Sigma^{+} \bar{K}^{0} \eta^{0} $ & $      0.30 \pm     0.09 $ \\
$\Sigma^{0} \pi^{+} \bar{K}^{0} $ & $     0.95 \pm     0.21 $ \\
$\Xi^{0} \pi^{0} \pi^{+} $ & $     1.5 \pm     0.3 $ \\
$\Xi^{0} \pi^{+} \eta^{0} $ & $     1.1 \pm     0.3 $ \\
$ \Xi^{0} K^{+} \bar{K}^{0} $ & $    0.34 \pm     0.06 $ \\
$\Xi^{-} \pi^{+} \pi^{+} $ & $     5.7 \pm     1.3 $ \\
$p \bar{K}^{0} \bar{K}^{0} $ & $     3.7 \pm     0.9 $ \\
$\Lambda^{0} \pi^{+} \bar{K}^{0} $ & $     3.8 \pm     0.9 $ \\
&\\
&\\
&\\
&\\
&\\
&\\
&\\
&\\
&\\
&\\
\hline
\end{tabular}
\begin{tabular}{|c|c|}
\hline
CS mode&$10^{3}{\cal B}$\\
\hline
$ \Sigma^{+} \pi^{0} \eta^{0} $ & $    5.3 \pm    1.1 $ \\
$ \Sigma^{+} \pi^{+} \pi^{-} $ & $    5.9 \pm    1.5 $ \\
$ \Sigma^{+} K^{0} \bar{K}^{0} $ & $    4.3 \pm     0.9 $ \\
$ \Sigma^{+} K^{+} K^{-} $ & $     0.58 \pm     0.09 $ \\
$\Sigma^{+} \eta^{0} \eta^{0}$ & $     0.31 \pm     0.09 $ \\
$ \Sigma^{0} \pi^{0} \pi^{+} $ & $    7.9 \pm    1.4 $ \\
$ \Sigma^{0} \pi^{+} \eta^{0} $ & $    5.1 \pm    1.1 $ \\
$ \Sigma^{0} K^{+} \bar{K}^{0} $ & $    1.5 \pm     0.5 $ \\
$\Sigma^{-} \pi^{+} \pi^{+} $ & $   16 \pm    3 $ \\
$ \Xi^{0} \pi^{0} K^{+} $ & $    3.6 \pm     0.8 $ \\
$ \Xi^{0} \pi^{+} K^{0}$ & $    8.4 \pm    2.5 $ \\
$ \Xi^{0} K^{+} \eta^{0} $ & $     0.42 \pm     0.13 $ \\
$p \pi^{0} \bar{K}^{0} $ & $   19 \pm    3 $ \\
$p \pi^{+} K^{-} $ & $   27 \pm    3 $ \\
$ n \pi^{+} \bar{K}^{0} $ & $    9.2 \pm    2.0 $ \\
$\Lambda^{0} \pi^{+} \eta^{0} $ & $   15 \pm    3 $ \\
$ \Lambda^{0} K^{+} \bar{K}^{0} $ & $    5.8 \pm     0.8 $ \\
&\\
&\\
&\\
\hline
\end{tabular}
\begin{tabular}{|c|c|}
\hline
DCS mode& $10^{4}{\cal B}$\\
\hline
$ \Sigma^{+} \pi^{0} K^{0} $ & $     2.6 \pm      0.2 $ \\
$ \Sigma^{+} \pi^{-} K^{+}  $ & $     1.4 \pm      0.3 $ \\
$ \Sigma^{+} K^{0} \eta^{0}  $ & $      0.020 \pm      0.014 $ \\
$ \Sigma^{0} \pi^{0} K^{+}  $ & $      0.076 \pm      0.059 $ \\
$ \Sigma^{0} \pi^{+} K^{0}  $ & $     2.5 \pm      0.2 $ \\
$ \Sigma^{0} K^{+} \eta^{0}  $ & $      0.010 \pm      0.007 $ \\
$ \Sigma^{-} \pi^{+} K^{+}  $ & $      1.3 \pm       0.1 $ \\
$ \Xi^{0} K^{0} K^{+}  $ & $      0.030 \pm       0.019 $ \\
$ \Xi^{-} K^{+} K^{+}  $ & $      0.057 \pm      0.032 $ \\
$ p \pi^{0} \pi^{0}  $ & $      7.2 \pm      1.8 $ \\
$ p \pi^{0} \eta^{0}  $ & $      11 \pm      2 $ \\
$ p \pi^{+} \pi^{-}  $ & $      14 \pm      4 $ \\
$ p K^{0} \bar{K}^{0}  $ & $     7.7 \pm     1.7 $ \\
$ p K^{+} K^{-}  $ & $     1.6 \pm     1.2 $ \\
$ p \eta^{0} \eta^{0}  $ & $      0.93 \pm      0.45 $ \\
$ n \pi^{+} \eta^{0}  $ & $     21 \pm      4 $ \\
$ n K^{+} \bar{K}^{0}  $ & $     16 \pm      3 $ \\
$ \Lambda^{0} \pi^{0} K^{+}  $ & $      5.0 \pm      1.0 $ \\
$ \Lambda^{0} \pi^{+} K^{0}  $ & $    9.7 \pm      2.0 $ \\
$ \Lambda^{0} K^{+} \eta^{0}  $ & $     0.90 \pm      0.22 $ \\

\hline
\end{tabular}
}
%\end{tabular}
\end{table}

\begin{table}
\caption{Numerical results for 
 ${\cal B}(\Xi_{c}^{0} \to {\bf B_n}MM')$.}\label{pre_Xic0}
  {\scriptsize
\begin{tabular}{|c|c|}
\hline
CF mode& $10^{2}{\cal B}$\\
\hline
$ \Sigma^{+} \pi^{0} K^{-} $ & $     7.9 \pm     1.4 $ \\
$\Sigma^{+} \pi^{-} \bar{K}^{0} $ & $    15 \pm     2 $ \\
$\Sigma^{+} K^{-} \eta^{0}$ & $     0.37 \pm     0.07 $ \\
$ \Sigma^{0} \pi^{0} \bar{K}^{0} $ & $     2.4 \pm     0.5 $ \\
$ \Sigma^{0} \pi^{+} K^{-} $ & $     3.9 \pm     1.0 $ \\
$\Sigma^{0} \bar{K}^{0} \eta^{0}$ & $     0.061 \pm     0.015 $ \\
$\Sigma^{-} \pi^{+} \bar{K}^{0}$ & $     0.34 \pm     0.05 $ \\
$ \Xi^{0} \pi^{0} \pi^{0} $ & $     7.2 \pm     1.5 $ \\
$ \Xi^{0} \pi^{0} \eta^{0} $ & $     1.0 \pm     0.1 $ \\
$\Xi^{0} \pi^{+} \pi^{-}$ & $    11 \pm     2 $ \\
$\Xi^{0} K^{0} \bar{K}^{0}$ & $     0.033 \pm     0.004 $ \\
$\Xi^{0} K^{+} K^{-} $ & $     0.30 \pm     0.03 $ \\
$\Xi^{0} \eta^{0} \eta^{0} $ & $(8.2 \pm 2.5)\times 10^{-4} $ \\
$\Xi^{-} \pi^{0} \pi^{+} $ & $     0.37 \pm     0.08 $ \\
$\Xi^{-} \pi^{+} \eta^{0} $ & $     0.93 \pm     0.11 $ \\
$\Xi^{-} K^{+} \bar{K}^{0} $ & $     0.077 \pm     0.014 $ \\
$ p K^{-} \bar{K}^{0} $ & $     1.7 \pm     0.3 $ \\
$n \bar{K}^{0} \bar{K}^{0} $ & $     0.77 \pm     0.09 $ \\
$ \Lambda^{0} \pi^{+} K^{-} $ & $     2.2 \pm     0.6 $ \\
$\Lambda^{0} \bar{K}^{0} \eta^{0} $ & $     0.057 \pm     0.024 $ \\
&\\
&\\
&\\
&\\
&\\
&\\
&\\
&\\
&\\
\hline
\end{tabular}
\begin{tabular}{|c|c|}
\hline
CS mode& $10^{3}{\cal B}$\\
\hline
$\Sigma^{+} \pi^{0} \pi^{-} $ & $    1.0 \pm     0.2 $ \\
$\Sigma^{+} \pi^{-} \eta^{0} $ & $    4.5 \pm     0.6 $ \\
$\Sigma^{+} K^{0} K^{-} $ & $    1.8 \pm     0.3 $ \\
$\Sigma^{0} \pi^{0} \pi^{0} $ & $    1.8 \pm     0.2 $ \\
$\Sigma^{0} \pi^{0} \eta^{0} $ & $    1.8 \pm     0.2 $ \\
$\Sigma^{0} K^{0} \bar{K}^{0} $ & $     0.039 \pm     0.007 $ \\
$\Sigma^{0} K^{+} K^{-} $ & $    1.2 \pm     0.2 $ \\
$\Sigma^{0} \eta^{0} \eta^{0} $ & $     0.39 \pm     0.01 $ \\
$\Sigma^{-} \pi^{0} \pi^{+} $ & $     1.0 \pm     0.2 $ \\
$\Sigma^{-} \pi^{+} \eta^{0} $ & $     0.44 \pm     0.13 $ \\
$\Sigma^{-} K^{+} \bar{K}^{0} $ & $     0.24 \pm     0.04 $ \\
$\Xi^{0} \pi^{0} K^{0} $ & $     0.59 \pm     0.28 $ \\
$\Xi^{0} \pi^{-} K^{+} $ & $    1.0 \pm     0.3 $ \\
$\Xi^{0} K^{0} \eta^{0} $ & $     0.013 \pm     0.005 $ \\
$\Xi^{-} \pi^{0} K^{+} $ & $     0.43 \pm     0.11 $ \\
$\Xi^{-} \pi^{+} K^{0} $ & $     0.61 \pm     0.10 $ \\
$\Xi^{-} K^{+} \eta^{0} $ & $     (2.3 \pm     0.8)\times10^{-3}$ \\
$p \pi^{0} K^{-} $ & $   13 \pm    2 $ \\
$p \pi^{-} \bar{K}^{0} $ & $   22 \pm    4 $ \\
$p K^{-} \eta^{0} $ & $    1.9 \pm     0.4 $ \\
$n \pi^{0} \bar{K}^{0} $ & $    2.4 \pm    1.2 $ \\
$n \pi^{+} K^{-} $ & $    8.9 \pm    2.8 $ \\
$n \bar{K}^{0} \eta^{0} $ & $     0.80 \pm     0.29 $ \\
$\Lambda^{0} \pi^{0} \pi^{0} $ & $    8.8\pm    1.1 $ \\
$\Lambda^{0} \pi^{0} \eta^{0} $ & $    2.0 \pm     0.3 $ \\
$\Lambda^{0} \pi^{+} \pi^{-} $ & $   17 \pm    2 $ \\
$\Lambda^{0} K^{0} \bar{K}^{0} $ & $     0.23 \pm     0.03 $ \\
$\Lambda^{0} K^{+} K^{-} $ & $     0.62 \pm     0.12 $ \\
$\Lambda^{0} \eta^{0} \eta^{0} $ & $     0.18 \pm     0.03 $ \\
\hline
\end{tabular}
\begin{tabular}{|c|c|}
\hline
DCS mode& $10^{5}{\cal B}$\\
\hline
$\Sigma^{+} \pi^{-} K^{0} $ & $    4.3 \pm     0.6 $ \\
$\Sigma^{0} \pi^{0} K^{0} $ & $    1.1 \pm     0.4 $ \\
$\Sigma^{0} \pi^{-} K^{+} $ & $    5.2 \pm     0.6 $ \\
$\Sigma^{0} K^{0} \eta^{0} $ & $     0.024\pm     0.016 $ \\
$\Sigma^{-} \pi^{0} K^{+} $ & $    5.1 \pm     0.6 $ \\
$\Sigma^{-} \pi^{+} K^{0} $ & $    3.1 \pm     0.4 $ \\
$\Sigma^{-} K^{+} \eta^{0} $ & $     0.047 \pm     0.031 $ \\
$\Xi^{0} K^{0} K^{0} $ & $     0.085 \pm    0.031 $ \\
$\Xi^{-} K^{0} K^{+} $ & $     0.040 \pm     0.015 $ \\
$p \pi^{-} \eta^{0} $ & $   64 \pm   10 $ \\
$p K^{0} K^{-} $ & $   43 \pm    5 $ \\
$n \pi^{0} \pi^{0} $ & $   16 \pm    2 $ \\
$n \pi^{0} \eta^{0} $ & $   32 \pm    5 $ \\
$n \pi^{+} \pi^{-} $ & $   31 \pm   4 $ \\
$n K^{0} \bar{K}^{0} $ & $    5.0 \pm     0.4 $ \\
$n K^{+} K^{-} $ & $   22 \pm    4 $ \\
$n \eta^{0} \eta^{0} $ & $     0.79 \pm     0.40 $ \\
$\Lambda^{0} \pi^{0} K^{0} $ & $    6.5 \pm    1.9 $ \\
$\Lambda^{0} \pi^{-} K^{+} $ & $   13 \pm    4 $ \\
$\Lambda^{0} K^{0} \eta^{0} $ & $     0.61 \pm     0.17 $ \\
&\\
&\\
&\\
&\\
&\\
&\\
&\\
&\\
&\\
\hline
\end{tabular}
}
\end{table}

Since $(a_i,b_i)$ and $(-a_i,b_i)$  give the same results in our $\chi^2$ fitting with $SU(3)_f$, 
both $\pm \langle \alpha \rangle$ are solutions which should be determined by experiments or dynamical models. 
We present the predictions for the up-down asymmetries of
$\langle \alpha \rangle (\Lambda_{c}^+,\Xi_{c}^{+},\Xi_{c}^{0} \to {\bf B_n}MM')$ 
in Tables~\ref{apre_Lcp}, \ref{apre_Xicp} and  \ref{apre_Xic0}, respectively,  by choosing $\langle \alpha \rangle(\Lambda_{c}^+ \to \Xi^- K^+ \pi^+)$ to be negative.  
One may also re-parametrize the real $SU(3)$ irreducible amplitudes with $a_i$ and $b_i$ in 
Eq.~(\ref{su3_amp})
 into  complex ones with $\tilde{a_i}=a_i+i\kappa(m_{23}^2)b_i$ and assume $\kappa({m_{23}^2})$ being the same for all modes in the
 $SU(3)_f$ limit. 
In this case, one more parameter can be reduced by considering the following transformations in $a_i$ and $b_i$ without changing the branching ratios,
given by 
\begin{eqnarray}
	a_{i}'&=&\cos(\theta)a_{i}+\kappa\sin(\theta)b_i\,, \nonumber\\
	\kappa b_{i}'&=&-\sin(\theta)a_{i}+\kappa\cos(\theta)b_i \,,
\end{eqnarray}
 which are equivalent to multiply an arbitrary overall phase $e^{i\theta}$ in the complex number parametrization, but  lose all information 
 about the up-down asymmetries~\cite{Geng:2018upx}.
Note that the fitting results of the branching ratios are slightly different from those in Ref.~\cite{Geng:2018upx} due to
 the kinematic  $\kappa(m_{23}^2)$ corrections. The  situation also occurs  in the semi-leptonic charmed baryon
 decays, indicating that $SU(3)_f$ is highly brokien in kinematics~\cite{Geng:2019bfz}.
\begin{table}
	\caption{Numerical results for 
		$\langle \alpha \rangle(\Lambda_{c}^{+} \to {\bf B_n}MM')$.}\label{apre_Lcp}
{\scriptsize
	\begin{tabular}{|c|c|}
		\hline
		CF mode& $\langle \alpha \rangle$ \\
		\hline
	$ \Lambda_{c}^{+} \to\Sigma^{+} \pi^{0} \pi^{0} $ & $       0.85 \pm       0.13 $ \\
	$ \Lambda_{c}^{+} \to\Sigma^{+} \pi^{0} \eta^{0} $ & $       0.81 \pm       0.18 $ \\
	$\Lambda_{c}^{+} \to\Sigma^{+} \pi^{+} \pi^{-} $ & $       0.16 \pm       0.27 $ \\
	$\Lambda_{c}^{+} \to \Sigma^{+} K^{0} \bar{K}^{0} $ & $       0.68 \pm       0.07 $ \\
	$\Lambda_{c}^{+} \to \Sigma^{+} K^{+} K^{-} $ & $      -0.06 \pm       0.11 $ \\
	$ \Lambda_{c}^{+} \to\Sigma^{+} \eta^{0} \eta^{0} $ & $       0.03 \pm       0.00 $ \\
	$ \Lambda_{c}^{+} \to\Sigma^{0} \pi^{0} \pi^{+} $ & $      -0.96 ^{+0.07}_{-0.04} $ \\
	$ \Lambda_{c}^{+} \to\Sigma^{0} \pi^{+} \eta^{0} $ & $       0.81 \pm       0.18 $ \\
	$\Lambda_{c}^{+} \to \Sigma^{0} K^{+} \bar{K}^{0} $ & $       0.30 \pm       0.60 $ \\
	$\Lambda_{c}^{+} \to\Sigma^{-} \pi^{+} \pi^{+} $ & $      -0.96 ^{+0.07 }_{-0.04}$ \\
	$ \Lambda_{c}^{+} \to\Xi^{0} \pi^{0} K^{+} $ & $       0.78 \pm       0.03 $ \\
	$\Lambda_{c}^{+} \to \Xi^{0} \pi^{+} K^{0} $ & $       0.96 \pm       0.00 $ \\
	$\Lambda_{c}^{+} \to \Xi^{-} \pi^{+} K^{+} $ & $      -0.78 \pm       0.13 $ \\
	$ \Lambda_{c}^{+} \to p \pi^{0} \bar{K}^{0} $ & $       0.11 \pm       0.28 $ \\
	$ \Lambda_{c}^{+} \to p \pi^{+} K^{-} $ & $       0.89 \pm       0.10 $ \\
	$\Lambda_{c}^{+} \to p \bar{K}^{0} \eta^{0} $ & $      -0.38 \pm       0.22 $ \\
	$\Lambda_{c}^{+} \to n \pi^{+} \bar{K}^{0} $ & $      -0.91 ^{+0.13}_{-0.09} $ \\
	$ \Lambda_{c}^{+} \to\Lambda^{0} \pi^{+} \eta^{0} $ & $       0.54 \pm       0.15 $ \\
	$\Lambda_{c}^{+} \to \Lambda^{0} K^{+} \bar{K}^{0} $ & $       0.41 \pm       0.08 $ \\
		\hline
	\end{tabular}
	\begin{tabular}{|c|c|}
		\hline
		CS mode& $\langle \alpha \rangle$\\
		\hline
	$\Lambda_{c}^{+} \to \Sigma^{+} \pi^{0} K^{0} $ & $       0.76 \pm       0.22 $ \\
	$\Lambda_{c}^{+} \to \Sigma^{+} \pi^{-} K^{+} $ & $       0.75 \pm       0.15 $ \\
	$\Lambda_{c}^{+} \to \Sigma^{+} K^{0} \eta^{0} $ & $      -0.05 \pm       0.07 $ \\
	$\Lambda_{c}^{+} \to \Sigma^{0} \pi^{0} K^{+} $ & $       0.75 \pm       0.10 $ \\
	$\Lambda_{c}^{+} \to \Sigma^{0} \pi^{+} K^{0} $ & $       0.75 \pm       0.22 $ \\
	$\Lambda_{c}^{+} \to \Sigma^{0} K^{+} \eta^{0} $ & $      -0.05 \pm       0.07 $ \\
	$\Lambda_{c}^{+} \to \Sigma^{-} \pi^{+} K^{+} $ & $       0.70 \pm       0.70 $ \\
	$\Lambda_{c}^{+} \to p \pi^{0} \pi^{0} $ & $      -0.95 \pm       0.05 $ \\
	$\Lambda_{c}^{+} \to p \pi^{0} \eta^{0} $ & $       0.84 \pm       0.09 $ \\
	$\Lambda_{c}^{+} \to p \pi^{+} \pi^{-} $ & $      -0.95 \pm       0.05 $ \\
	$\Lambda_{c}^{+} \to p K^{0} \bar{K}^{0} $ & $       0.84 \pm       0.05 $ \\
	$\Lambda_{c}^{+} \to p K^{+} K^{-} $ & $      -0.91 \pm       0.09 $ \\
	$\Lambda_{c}^{+} \to p \eta^{0} \eta^{0} $ & $       0.62 \pm       0.21 $ \\
	$\Lambda_{c}^{+} \to n \pi^{+} \eta^{0} $ & $       0.85 \pm       0.09 $ \\
	$\Lambda_{c}^{+} \to n K^{+} \bar{K}^{0} $ & $       0.94 \pm       0.03 $ \\
	$\Lambda_{c}^{+} \to \Lambda^{0} \pi^{0} K^{+} $ & $       0.97 \pm       0.00 $ \\
	$\Lambda_{c}^{+} \to \Lambda^{0} \pi^{+} K^{0} $ & $       0.97 \pm       0.00 $ \\
	$\Lambda_{c}^{+} \to \Lambda^{0} K^{+} \eta^{0} $ & $      -0.28 \pm       0.28 $ \\
	&\\
		\hline
	\end{tabular}
	\begin{tabular}{|c|c|}
		\hline
		DCS mode& $\langle \alpha \rangle$\\
		\hline
		$\Lambda_{c}^{+} \to\Sigma^{+} K^{0} K^{0} $ & $      -0.43 \pm       0.32 $ \\
		$\Lambda_{c}^{+} \to\Sigma^{0} K^{0} K^{+} $ & $      -0.43 \pm       0.32 $ \\
		$\Lambda_{c}^{+} \to \Sigma^{-} K^{+} K^{+} $ & $      -0.43 \pm       0.31 $ \\
		$\Lambda_{c}^{+} \to p \pi^{0} K^{0} $ & $       0.93^{+0.07}_{-0.10} $ \\
		$\Lambda_{c}^{+} \to p \pi^{-} K^{+} $ & $       0.93 ^{+0.07}_{-0.10} $ \\
		$\Lambda_{c}^{+} \to p K^{0} \eta^{0} $ & $      -0.38 \pm       0.45 $ \\
		$\Lambda_{c}^{+} \to n \pi^{0} K^{+} $ & $       0.93^{+0.07}_{-0.10} $ \\
		$\Lambda_{c}^{+} \to n \pi^{+} K^{0} $ & $       0.93^{+0.07}_{-0.10} $ \\
		$\Lambda_{c}^{+} \to n K^{+} \eta^{0} $ & $      -0.38 \pm       0.45 $ \\	
		&\\	
		&\\	
		&\\	
		&\\	
		&\\	
		&\\	
		&\\	
		&\\	
		&\\	
		&\\	
		\hline	
	\end{tabular}
	}
\end{table}

\begin{table}
	\caption{Numerical results for 
		 $\langle \alpha \rangle(\Xi_{c}^{+} \to {\bf B_n}MM')$.}\label{apre_Xicp}
{\scriptsize 
	\begin{tabular}{|c|c|}
		\hline
		CF mode& $\langle \alpha \rangle$\\
		\hline
		$\Xi_{c}^{+} \to  \Sigma^{+} \pi^{0} \bar{K}^{0} $ & $       0.67 \pm       0.22 $ \\
		$\Xi_{c}^{+} \to  \Sigma^{+} \pi^{+} K^{-} $ & $       0.86 ^{+0.14}_{-0.15} $ \\
		$\Xi_{c}^{+} \to  \Sigma^{+} \bar{K}^{0} \eta^{0} $ & $       0.21 \pm       0.10 $ \\
		$\Xi_{c}^{+} \to  \Sigma^{0} \pi^{+} \bar{K}^{0} $ & $      -0.81 ^{+0.36}_{-0.19} $ \\
		$\Xi_{c}^{+} \to  \Xi^{0} \pi^{0} \pi^{+} $ & $      -0.81 ^{+0.36}_{-0.19}$ \\
		$\Xi_{c}^{+} \to  \Xi^{0} \pi^{+} \eta^{0} $ & $       0.96 \pm       0.04 $ \\
		$\Xi_{c}^{+} \to  \Xi^{0} K^{+} \bar{K}^{0} $ & $       0.54 \pm       0.17 $ \\
		$\Xi_{c}^{+} \to \Xi^{-} \pi^{+} \pi^{+} $ & $      -0.81 ^{+0.37}_{-0.19}$ \\
		$\Xi_{c}^{+} \to  p \bar{K}^{0} \bar{K}^{0} $ & $      -0.87^{+0.31}_{-0.13} $ \\
		$\Xi_{c}^{+} \to  \Lambda^{0} \pi^{+} \bar{K}^{0} $ & $      -0.86^{+0.33}_{-0.14} $ \\
		&\\	
		&\\	
		&\\	
		&\\	
		&\\	
		&\\	
		&\\	
		&\\	
		&\\	
		&\\	
		\hline
	\end{tabular}
	\begin{tabular}{|c|c|}
		\hline
		CS mode& $\langle \alpha \rangle$\\
		\hline
		$\Xi_{c}^{+} \to  \Sigma^{+} \pi^{0} \pi^{0} $ & $      -0.95 ^{+0.65}_{-0.05} $ \\
		$\Xi_{c}^{+} \to  \Sigma^{+} \pi^{0} \eta^{0} $ & $       0.33 \pm       0.19 $ \\
		$\Xi_{c}^{+} \to  \Sigma^{+} \pi^{+} \pi^{-} $ & $      -0.96 \pm       0.04 $ \\
		$\Xi_{c}^{+} \to  \Sigma^{+} K^{0} \bar{K}^{0} $ & $       0.71 \pm       0.06 $ \\
		$\Xi_{c}^{+} \to  \Sigma^{+} K^{+} K^{-} $ & $      -0.87 \pm       0.11 $ \\
		$\Xi_{c}^{+} \to  \Sigma^{+} \eta^{0} \eta^{0} $ & $       0.63 \pm       0.09 $ \\
		$\Xi_{c}^{+} \to  \Sigma^{0} \pi^{0} \pi^{+} $ & $      -0.97 \pm       0.03 $ \\
		$\Xi_{c}^{+} \to  \Sigma^{0} \pi^{+} \eta^{0} $ & $       0.34 \pm       0.19 $ \\
		$\Xi_{c}^{+} \to  \Sigma^{0} K^{+} \bar{K}^{0} $ & $       0.96 \pm       0.01 $ \\
		$\Xi_{c}^{+} \to \Sigma^{-} \pi^{+} \pi^{+} $ & $      -0.96 \pm       0.03 $ \\
		$\Xi_{c}^{+} \to  \Xi^{0} \pi^{0} K^{+} $ & $       0.95 \pm       0.03 $ \\
		$\Xi_{c}^{+} \to  \Xi^{0} \pi^{+} K^{0} $ & $       0.96 \pm       0.01 $ \\
		$\Xi_{c}^{+} \to  \Xi^{0} K^{+} \eta^{0} $ & $       0.85 \pm       0.08 $ \\
		$\Xi_{c}^{+} \to  \Xi^{-} \pi^{+} K^{+} $ & $       0.70^{+0.30}_{-0.70} $ \\
		$\Xi_{c}^{+} \to  p \pi^{0} \bar{K}^{0} $ & $       0.30 \pm       0.17 $ \\
		$\Xi_{c}^{+} \to  p \pi^{+} K^{-} $ & $       0.94 ^{+0.06}_{-0.07} $ \\
		$\Xi_{c}^{+} \to  p \bar{K}^{0} \eta^{0} $ & $       0.49 \pm       0.28 $ \\
		$\Xi_{c}^{+} \to  n \pi^{+} \bar{K}^{0} $ & $      -0.97 \pm       0.02 $ \\
		$\Xi_{c}^{+} \to  \Lambda^{0} \pi^{+} \eta^{0} $ & $       0.96 \pm       0.02 $ \\
		$\Xi_{c}^{+} \to \Lambda^{0} K^{+} \bar{K}^{0} $ & $       0.91 \pm       0.06 $ \\
		\hline
	\end{tabular}
	\begin{tabular}{|c|c|}
		\hline
		DCS mode& $\langle \alpha \rangle$\\
		\hline
		$\Xi_{c}^{+} \to  \Sigma^{+} \pi^{0} K^{0} $ & $      -0.26 \pm       0.15 $ \\
		$\Xi_{c}^{+} \to  \Sigma^{+} \pi^{-} K^{+} $ & $       0.80 \pm       0.13 $ \\
		$\Xi_{c}^{+} \to  \Sigma^{+} K^{0} \eta^{0} $ & $       0.61 \pm       0.33 $ \\
		$\Xi_{c}^{+} \to  \Sigma^{0} \pi^{0} K^{+} $ & $      -0.05 \pm       0.21 $ \\
		$\Xi_{c}^{+} \to \Sigma^{0} \pi^{+} K^{0} $ & $      -0.26 \pm       0.15 $ \\
		$\Xi_{c}^{+} \to  \Sigma^{0} K^{+} \eta^{0} $ & $       0.60 \pm       0.33 $ \\
		$\Xi_{c}^{+} \to  \Sigma^{-} \pi^{+} K^{+} $ & $      -0.94 ^{+0.07}_{-0.06} $ \\
		$\Xi_{c}^{+} \to \Xi^{0} K^{0} K^{+} $ & $      -0.84  ^{+0.18}_{-0.16} $ \\
		$\Xi_{c}^{+} \to  \Xi^{-} K^{+} K^{+} $ & $      -0.83^{+ 0.18}_{-0.17} $ \\
		$\Xi_{c}^{+} \to  p \pi^{0} \pi^{0} $ & $      -0.22 \pm       0.38 $ \\
		$\Xi_{c}^{+} \to  p \pi^{0} \eta^{0} $ & $       0.97 \pm       0.00 $ \\
		$\Xi_{c}^{+} \to  p \pi^{+} \pi^{-} $ & $      -0.22 \pm       0.38 $ \\
		$\Xi_{c}^{+} \to  p K^{0} \bar{K}^{0} $ & $       0.96 \pm       0.02 $ \\
		$\Xi_{c}^{+} \to  p K^{+} K^{-} $ & $       0.14 \pm       0.24 $ \\
		$\Xi_{c}^{+} \to  p \eta^{0} \eta^{0} $ & $       0.93 ^{+0.07}_{-0.12} $ \\
		$\Xi_{c}^{+} \to  n \pi^{+} \eta^{0} $ & $       0.97 \pm       0.00 $ \\
		$\Xi_{c}^{+} \to  n K^{+} \bar{K}^{0} $ & $       0.83 \pm       0.06 $ \\
		$\Xi_{c}^{+} \to  \Lambda^{0} \pi^{0} K^{+} $ & $       0.80 \pm       0.12 $ \\
		$\Xi_{c}^{+} \to  \Lambda^{0} \pi^{+} K^{0} $ & $       0.80 \pm       0.12 $ \\
		$\Xi_{c}^{+} \to  \Lambda^{0} K^{+} \eta^{0} $ & $      -0.03 \pm       0.35 $ \\
		\hline
	\end{tabular}
	}
\end{table}

\begin{table}
	\caption{Numerical results for 
		 $\langle \alpha \rangle(\Xi_{c}^{0} \to {\bf B_n}MM')$.}\label{apre_Xic0}
{\scriptsize
	\begin{tabular}{|c|c|}
		\hline
		CF mode& $\langle \alpha \rangle$\\
		\hline
		$\Xi_{c}^{0} \to  \Sigma^{+} \pi^{0} K^{-} $ & $       0.94 \pm       0.02 $ \\
		$\Xi_{c}^{0} \to  \Sigma^{+} \pi^{-} \bar{K}^{0} $ & $       0.89 \pm       0.05 $ \\
		$\Xi_{c}^{0} \to  \Sigma^{+} K^{-} \eta^{0} $ & $       0.79 \pm       0.03 $ \\
		$\Xi_{c}^{0} \to  \Sigma^{0} \pi^{0} \bar{K}^{0} $ & $       0.44 \pm       0.17 $ \\
		$\Xi_{c}^{0} \to  \Sigma^{0} \pi^{+} K^{-} $ & $       0.95 \pm       0.03 $ \\
		$\Xi_{c}^{0} \to  \Sigma^{0} \bar{K}^{0} \eta^{0} $ & $       0.96 \pm       0.01 $ \\
		$\Xi_{c}^{0} \to  \Sigma^{-} \pi^{+} \bar{K}^{0} $ & $      -0.96^{+0.06}_{-0.04} $ \\
		$\Xi_{c}^{0} \to  \Xi^{0} \pi^{0} \pi^{0} $ & $       0.86 \pm       0.05 $ \\
		$\Xi_{c}^{0} \to  \Xi^{0} \pi^{0} \eta^{0} $ & $       0.42 \pm       0.18 $ \\
		$\Xi_{c}^{0} \to  \Xi^{0} \pi^{+} \pi^{-} $ & $       0.97 \pm       0.01 $ \\
		$\Xi_{c}^{0} \to  \Xi^{0} K^{0} \bar{K}^{0} $ & $       0.32 \pm       0.52 $ \\
		$\Xi_{c}^{0} \to \Xi^{0} K^{+} K^{-} $ & $      -0.07 \pm       0.13 $ \\
		$\Xi_{c}^{0} \to  \Xi^{0} \eta^{0} \eta^{0} $ & $      -0.18 \pm       0.83 $ \\
		$\Xi_{c}^{0} \to  \Xi^{-} \pi^{0} \pi^{+} $ & $      -0.81 ^{+ 0.37}_{-0.19} $ \\
		$\Xi_{c}^{0} \to  \Xi^{-} \pi^{+} \eta^{0} $ & $      -0.09 \pm       0.11 $ \\
		$\Xi_{c}^{0} \to  \Xi^{-} K^{+} \bar{K}^{0} $ & $       0.47 \pm       0.12 $ \\
		$\Xi_{c}^{0} \to  p K^{-} \bar{K}^{0} $ & $      -0.96 \pm       0.03 $ \\
		$\Xi_{c}^{0} \to  n \bar{K}^{0} \bar{K}^{0} $ & $      -0.93^{+ 0.11}_{-0.07} $ \\
		$\Xi_{c}^{0} \to  \Lambda^{0} \pi^{0} \bar{K}^{0} $ & $       0.01 \pm       0.62 $ \\
		$\Xi_{c}^{0} \to  \Lambda^{0} \pi^{+} K^{-} $ & $      -0.91 \pm       0.08 $ \\
		$\Xi_{c}^{0} \to \Lambda^{0} \bar{K}^{0} \eta^{0} $ & $      -0.74 \pm       0.22 $ \\
		&\\
		&\\
		&\\
		&\\
		&\\
		&\\
		&\\
		&\\
		&\\
		\hline
	\end{tabular}
	\begin{tabular}{|c|c|}
		\hline
		CS mode&$\langle \alpha \rangle$\\
		\hline
		$\Xi_{c}^{0} \to  \Sigma^{+} \pi^{0} \pi^{-} $ & $      -0.97 \pm       0.03 $ \\
		$\Xi_{c}^{0} \to  \Sigma^{+} \pi^{-} \eta^{0} $ & $       0.73 \pm       0.10 $ \\
		$\Xi_{c}^{0} \to  \Sigma^{+} K^{0} K^{-} $ & $       0.82 \pm       0.03 $ \\
		$\Xi_{c}^{0} \to  \Sigma^{0} \pi^{0} \pi^{0} $ & $      -0.96 \pm       0.02 $ \\
		$\Xi_{c}^{0} \to  \Sigma^{0} \pi^{0} \eta^{0} $ & $       0.89 \pm       0.06 $ \\
		$\Xi_{c}^{0} \to  \Sigma^{0} \pi^{+} \pi^{-} $ & $      -0.95 ^{+ 0.64}_{-0.05} $ \\
		$\Xi_{c}^{0} \to  \Sigma^{0} K^{0} \bar{K}^{0} $ & $       0.95 \pm       0.05 $ \\
		$\Xi_{c}^{0} \to  \Sigma^{0} K^{+} K^{-} $ & $       0.54 \pm       0.02 $ \\
		$\Xi_{c}^{0} \to  \Sigma^{0} \eta^{0} \eta^{0} $ & $       0.63 \pm       0.09 $ \\
		$\Xi_{c}^{0} \to  \Sigma^{-} \pi^{0} \pi^{+} $ & $      -0.97 \pm       0.03 $ \\
		$\Xi_{c}^{0} \to  \Sigma^{-} \pi^{+} \eta^{0} $ & $       0.78 \pm       0.10 $ \\
		$\Xi_{c}^{0} \to  \Sigma^{-} K^{+} \bar{K}^{0} $ & $       0.07 \pm       0.16 $ \\
		$\Xi_{c}^{0} \to  \Xi^{0} \pi^{0} K^{0} $ & $       0.37 \pm       0.32 $ \\
		$\Xi_{c}^{0} \to \Xi^{0} \pi^{-} K^{+} $ & $       0.16 \pm       0.33 $ \\
		$\Xi_{c}^{0} \to  \Xi^{0} K^{0} \eta^{0} $ & $       0.16 \pm       0.15 $ \\
		$\Xi_{c}^{0} \to  \Xi^{-} \pi^{0} K^{+} $ & $       0.18 \pm       0.15 $ \\
		$\Xi_{c}^{0} \to \Xi^{-} \pi^{+} K^{0} $ & $       0.08 \pm       0.19 $ \\
		$\Xi_{c}^{0} \to  \Xi^{-} K^{+} \eta^{0} $ & $      -0.83 ^{+0.21}_{-0.17} $ \\
		$\Xi_{c}^{0} \to  p \pi^{0} K^{-} $ & $       0.75 \pm       0.07 $ \\
		$\Xi_{c}^{0} \to  p \pi^{-} \bar{K}^{0} $ & $       0.97 \pm       0.00 $ \\
		$\Xi_{c}^{0} \to  p K^{-} \eta^{0} $ & $       0.04 \pm       0.14 $ \\
		$\Xi_{c}^{0} \to  n \pi^{0} \bar{K}^{0} $ & $       0.88 ^{+ 0.12}_{-0.14} $ \\
		$\Xi_{c}^{0} \to  n \pi^{+} K^{-} $ & $       0.10 \pm       0.20 $ \\
		$\Xi_{c}^{0} \to  n \bar{K}^{0} \eta^{0} $ & $      -0.23 \pm       0.33 $ \\
		$\Xi_{c}^{0} \to  \Lambda^{0} \pi^{0} \pi^{0} $ & $       0.82 \pm       0.04 $ \\
$\Xi_{c}^{0} \to  \Lambda^{0} \pi^{0} \eta^{0} $ & $       0.96 \pm       0.02 $ \\
		$\Xi_{c}^{0} \to  \Lambda^{0} \pi^{+} \pi^{-} $ & $       0.82 \pm       0.04 $ \\
		$\Xi_{c}^{0} \to  \Lambda^{0} K^{0} \bar{K}^{0} $ & $       0.92 ^{+0.08}_{- 0.09} $ \\
		$ \Lambda^{0} K^{+} K^{-} $ & $       0.71 \pm       0.15 $ \\
		$\Xi_{c}^{0} \to  \Lambda^{0} \eta^{0} \eta^{0} $ & $      -0.90 \pm       0.10 $ \\
		\hline
	\end{tabular}
	\begin{tabular}{|c|c|}
		\hline
		DCS mode& $\langle \alpha \rangle$\\
		\hline
		$\Xi_{c}^{0} \to  \Sigma^{+} \pi^{-} K^{0} $ & $      -0.94 ^{+0.07}_{-0.06} $ \\
		$\Xi_{c}^{0} \to  \Sigma^{0} \pi^{0} K^{0} $ & $      -0.05 \pm       0.21 $ \\
		$\Xi_{c}^{0} \to  \Sigma^{0} \pi^{-} K^{+} $ & $      -0.26 \pm       0.15 $ \\
		$\Xi_{c}^{0} \to  \Sigma^{0} K^{0} \eta^{0} $ & $       0.60 \pm       0.33 $ \\
		$\Xi_{c}^{0} \to  \Sigma^{-} \pi^{0} K^{+} $ & $      -0.26 \pm       0.15 $ \\
		$\Xi_{c}^{0} \to  \Sigma^{-} \pi^{+} K^{0} $ & $       0.80 \pm       0.13 $ \\
		$\Xi_{c}^{0} \to  \Sigma^{-} K^{+} \eta^{0} $ & $       0.60 \pm       0.33 $ \\
		$\Xi_{c}^{0} \to  \Xi^{0} K^{0} K^{0} $ & $      -0.84^{ +0.18}_{-0.16} $ \\
		$\Xi_{c}^{0} \to  \Xi^{-} K^{0} K^{+} $ & $      -0.83 ^{+0.18}_{-0.17} $ \\
		$\Xi_{c}^{0} \to  p \pi^{-} \eta^{0} $ & $       0.97 \pm       0.00 $ \\
		$\Xi_{c}^{0} \to  p K^{0} K^{-} $ & $       0.83 \pm       0.06 $ \\
		$\Xi_{c}^{0} \to  n \pi^{0} \pi^{0} $ & $      -0.22 \pm       0.38 $ \\
		$\Xi_{c}^{0} \to  n \pi^{0} \eta^{0} $ & $       0.97 \pm       0.00 $ \\
		$\Xi_{c}^{0} \to  n \pi^{+} \pi^{-} $ & $      -0.22 \pm       0.38 $ \\
		$\Xi_{c}^{0} \to  n K^{0} \bar{K}^{0} $ & $       0.14 \pm       0.24 $ \\
		$\Xi_{c}^{0} \to  n K^{+} K^{-} $ & $       0.96 \pm       0.02 $ \\
		$\Xi_{c}^{0} \to  n \eta^{0} \eta^{0} $ & $       0.93 ^{+ 0.12}_{-0.07} $ \\
		$\Xi_{c}^{0} \to  \Lambda^{0} \pi^{0} K^{0} $ & $       0.80 \pm       0.12 $ \\
		$\Xi_{c}^{0} \to  \Lambda^{0} \pi^{-} K^{+} $ & $       0.80 \pm       0.12 $ \\
		$\Xi_{c}^{0} \to  \Lambda^{0} K^{0} \eta^{0} $ & $      -0.03 \pm       0.35 $ \\
		&\\
		&\\
		&\\
		&\\
		&\\
		&\\
		&\\
		&\\
		&\\
		&\\
		\hline
	\end{tabular}
	}
\end{table}

 We can also calculate the up-down asymmetries for the decays with the final states involving the physical $ K_{S}^0$ and $K_{L}^0$ particles,
  where $K_{S}^{0}=\frac{1}{\sqrt{2}}(K^{0}+\bar{K}^{0}) $ and $K_{L}^{0}=\frac{1}{\sqrt{2}}(K^{0}-\bar{K}^{0}) $ with ignoring CP violation. 
  The numerical values for the decay branching ratios and up-down asymmetries are presented in Table~\ref{klks}. 
 It is interesting to note that $\langle \alpha \rangle( \Lambda_{c}^{+} \to \Sigma^{+} K_{S}^{0} K_{L}^{0}) = -0.44 \pm 0.32 $ and $\langle \alpha \rangle( \Xi_{c}^{0} \to \Xi^{0} K_{S}^{0} K_{L}^{0}) = -0.85 ^{+0.17}_{-0.15} $ 
  are the same as $\langle \alpha \rangle( \Lambda_{c}^{+} \to \Sigma^{+} K^{0} K^{0})$ and $\langle \alpha \rangle( \Xi_{c}^{0} \to \Xi^{0} K^{0} K^{0}) $,
  in which the former two modes are dominated by the CF processes, whereas the later two the DCS ones.
 Clearly, these two modes can be used to test the s-wave dominance  assumption for the meson-pairs in the decays . 
\begin{table}
	\caption{Decay branching ratios and averaged up-down asymmetries for CF and DCS mixed processes involving $K_{S}^{0}$ and $K_{L}^{0}$.}\label{klks}
	{\scriptsize
	\begin{tabular}{cc}
		\begin{tabular}[t]{|c|c|c|}
			\hline
			Channel  & ${\cal B}$ &$\langle \alpha \rangle$\\
			\hline
			$ \Lambda_{c}^{+} \to \Sigma^{0} K^{+} K_{S}^{0} $ & $(1.44\pm 0.52)\times 10^{-4}$ &$       0.42 \pm       0.55 $ \\
			$ \Lambda_{c}^{+} \to \Sigma^{0} K^{+} K_{L}^{0} $ &$(1.47\pm 0.51)\times 10^{-4}$ &$       0.16 \pm       0.61 $ \\
			$ \Lambda_{c}^{+} \to p \pi^{0} K_{S}^{0} $ &$(1.26\pm 0.21)\times 10^{-2}$ &$       0.16 \pm       0.28 $ \\
			$ \Lambda_{c}^{+} \to p \pi^{0} K_{L}^{0} $ & $(1.09\pm 0.19)\times 10^{-2}$&$       0.06 \pm       0.29 $ \\
			$ \Lambda_{c}^{+} \to p \eta^{0} K_{S}^{0} $ &$(3.40\pm 0.54)\times 10^{-3}$& $      -0.35 \pm       0.21 $ \\
			$ \Lambda_{c}^{+} \to p \eta^{0} K_{L}^{0} $ &$(3.52\pm 0.56)\times 10^{-3}$ &$      -0.42 \pm       0.23 $ \\
			$ \Lambda_{c}^{+} \to n \pi^{+} K_{S}^{0} $ &$(5.80\pm 0.72)\times 10^{-3}$ &$      -0.96 ^{+0.06}_{-0.04} $ \\
			$ \Lambda_{c}^{+} \to n \pi^{+} K_{L}^{0} $ &$(5.88\pm 0.77)\times 10^{-3}$& $      -0.83 ^{+0.20}_{-0.17} $ \\
			\hline
			$ \Lambda_{c}^{+} \to \Sigma^{+} K_{S}^{0} K_{S}^{0} $ &$(1.77\pm 0.42)\times 10^{-3}$ &$       0.69 \pm       0.07 $ \\
			$ \Lambda_{c}^{+} \to \Sigma^{+} K_{S}^{0} K_{L}^{0} $ &$(7.56\pm 2.94)\times 10^{-7}$ &$      -0.44 \pm       0.32 $ \\
			$ \Lambda_{c}^{+} \to \Sigma^{+} K_{L}^{0} K_{L}^{0} $ &$(1.68\pm 0.41)\times 10^{-3}$ &$       0.71 \pm       0.07 $ \\
			\hline
			$ \Xi_{c}^{+} \to \Sigma^{+} \pi^{0} K_{S}^{0} $ &$(1.10\pm 0.66)\times 10^{-2}$& $       0.79 \pm       0.20 $ \\
			$ \Xi_{c}^{+} \to \Sigma^{+} \pi^{0} K_{L}^{0} $ & $(1.00\pm 0.70)\times 10^{-2}$&$       0.52 \pm       0.23 $ \\
			$ \Xi_{c}^{+} \to \Sigma^{+} \eta^{0} K_{S}^{0} $ &$(1.60\pm 0.45)\times 10^{-3}$& $       0.22 \pm       0.10 $ \\
			$ \Xi_{c}^{+} \to \Sigma^{+} \eta^{0} K_{L}^{0} $ &$(1.45\pm 0.47)\times 10^{-3}$ &$       0.20 \pm       0.11 $ \\
			$ \Xi_{c}^{+} \to \Sigma^{0} \pi^{+} K_{S}^{0} $ & $(3.60\pm 0.90)\times 10^{-3}$&$      -0.87 ^{+0.33}_{-0.13} $ \\
			$ \Xi_{c}^{+} \to \Sigma^{0} \pi^{+} K_{L}^{0} $ &$(6.20\pm 0.13)\times 10^{-3}$& $      -0.76^{+0.37}_{-0.24} $ \\
			$ \Xi_{c}^{+} \to \Xi^{0} K^{+} K_{S}^{0} $ &$(1.85\pm 0.34)\times 10^{-3}$& $       0.52 \pm       0.17 $ \\
			$ \Xi_{c}^{+} \to \Xi^{0} K^{+} K_{L}^{0} $ &$(1.75\pm 0.34)\times 10^{-3}$& $       0.57 \pm       0.17 $ \\
			$ \Xi_{c}^{+} \to \Lambda^{0} \pi^{+} K_{S}^{0} $&$(1.94\pm 0.43)\times 10^{-2}$ & $      -0.73^{+0.47}_{-0.23} $ \\
			$ \Xi_{c}^{+} \to \Lambda^{0} \pi^{+} K_{L}^{0} $&$(1.99\pm 0.49)\times 10^{-2}$ & $      -0.94 ^{ 0.17}_{0.06} $ \\
			\hline
			$ \Xi_{c}^{+} \to p K_{S}^{0} K_{S}^{0} $ &$(1.06\pm 0.23)\times 10^{-2}$ &$      -0.65 ^{+0.50}_{-0.35} $ \\
			$ \Xi_{c}^{+} \to p K_{S}^{0} K_{L}^{0} $ & $(1.92\pm 0.44)\times 10^{-2}$ &$      -0.88^{+ 0.31}_{-0.12} $ \\
			$ \Xi_{c}^{+} \to p K_{L}^{0} K_{L}^{0} $ &$(9.36\pm 3.07)\times 10^{-3}$ & $      -0.97 \pm       0.01 $ \\
			\hline
		\end{tabular}
	&
		\begin{tabular}[t]{|c|c|c|}
			\hline
			Channel  & ${\cal B}$ &$\langle \alpha \rangle$\\
			\hline
			$ \Xi_{c}^{0} \to \Sigma^{+} \pi^{-} K_{S}^{0} $ & $(7.43\pm 1.01)\times 10^{-2}$ &$       0.90 \pm       0.05 $ \\
			$ \Xi_{c}^{0} \to \Sigma^{+} \pi^{-} K_{L}^{0} $ &$(7.48\pm 1.01)\times 10^{-2}$  &$       0.87 \pm       0.06 $ \\
			$ \Xi_{c}^{0} \to \Sigma^{0} \pi^{0} K_{S}^{0} $ &$(1.19\pm 0.25)\times 10^{-2}$  &$       0.44 \pm       0.17 $ \\
			$ \Xi_{c}^{0} \to \Sigma^{0} \pi^{0} K_{L}^{0} $ &$(1.29\pm 0.25)\times 10^{-2}$  &$       0.43 \pm       0.17 $ \\
			$ \Xi_{c}^{0} \to \Sigma^{-} \pi^{+} K_{S}^{0} $ &$(1.84\pm 0.29)\times 10^{-3}$  &$      -0.97 \pm       0.02 $ \\
			$ \Xi_{c}^{0} \to \Sigma^{-} \pi^{+} K_{L}^{0} $ &$(1.69\pm 0.23)\times 10^{-3}$ &$      -0.92 ^{+0.14}_{-0.08} $ \\
			$ \Xi_{c}^{0} \to \Xi^{-} K^{+} K_{S}^{0} $ & $(4.20\pm 0.73)\times 10^{-4}$ &$       0.45 \pm       0.12 $ \\
			$ \Xi_{c}^{0} \to \Xi^{-} K^{+} K_{L}^{0} $ &$(3.93\pm 0.73)\times 10^{-4}$& $       0.51 \pm       0.12 $ \\
			$ \Xi_{c}^{0} \to p K^{-} K_{S}^{0} $ &$(8.18\pm 1.13)\times 10^{-3}$& $      -0.89^{+ 0.12}_{-0.11} $ \\
			$ \Xi_{c}^{0} \to p K^{-} K_{L}^{0} $ &$(9.42\pm 2.00)\times 10^{-3}$& $      -0.95 \pm       0.05 $ \\
	%		$ \Xi_{c}^{0} \to \Lambda^{0} \pi^{0} K_{S}^{0} $ &$(1.56\pm 1.65)\times 10^{-3}$ &$       0.14 \pm       0.52 $ \\
		%	$ \Xi_{c}^{0} \to \Lambda^{0} \pi^{0} K_{L}^{0} $ &$(0.87\pm 1.09)\times 10^{-3}$ &$      -0.17 \pm       0.75 $ \\
			$ \Xi_{c}^{0} \to \Lambda^{0} \eta^{0} K_{S}^{0} $ &$(2.69\pm 1.16)\times 10^{-4}$& $      -0.61 \pm       0.28 $ \\
			$ \Xi_{c}^{0} \to \Lambda^{0} \eta^{0} K_{L}^{0} $ &$(3.23\pm 1.27)\times 10^{-4}$& $      -0.84 \pm       0.16 $ \\
			\hline
			$ \Xi_{c}^{0} \to \Xi^{0} K_{S}^{0} K_{S}^{0} $ &$(1.85\pm 0.21)\times 10^{-4}$ &$       0.37 \pm       0.49 $ \\
			$ \Xi_{c}^{0} \to \Xi^{0} K_{S}^{0} K_{L}^{0}$ &$(4.70\pm 1.71)\times 10^{-7}$ &$      -0.85 ^{+0.17}_{-0.15} $ \\
			$ \Xi_{c}^{0} \to \Xi^{0} K_{L}^{0} K_{L}^{0} $ & $(1.95\pm 0.19)\times 10^{-4}$&$       0.25 \pm       0.53 $ \\
			$ \Xi_{c}^{0} \to n K_{S}^{0} K_{S}^{0} $ &$(1.66\pm 0.23)\times 10^{-3}$& $      -0.96 \pm       0.04 $ \\
			$ \Xi_{c}^{0} \to n K_{S}^{0} K_{L}^{0} $ &$(3.97\pm 0.49)\times 10^{-3}$ &$      -0.93 ^{+0.12}_{-0.07} $ \\
			$ \Xi_{c}^{0} \to n K_{L}^{0} K_{L}^{0} $ &$(2.35\pm 0.27)\times 10^{-3}$ &$      -0.88 ^{+ 0.16}_{-0.14} $ \\
			&&\\
			&&\\
			&&\\
			&&\\
			&&\\
			&&\\
			\hline
		\end{tabular}
\end{tabular}		
}
\end{table}

\section{Discussions and Conclusions}
We have studied the up-down asymmetries in the three-body anti-triplet ${\bf B_c}\to {\bf B_n}MM'$ decays
in the approach of the $SU(3)_f$ symmetry.
In our analysis, we have only concentrated on the s-wave $MM'$-pair contributions, 
so that the decays  only depend on 12 real irreducible parity conserving and violating  amplitudes.
% correspond to s-wave and p-wave decay respectively.
%
With the minimum $\chi^2$ fit to the 16 data points,  we have obtained a fit with $\chi^2/d.o.f=2.4$, 
which is not relatively good but it will be reduced when more decay branching ratios or up-down asymmetries of the three-body modes are measured in the future.
The predictions of the decay branching ratios are slightly
different from those in Ref.~\cite{Geng:2018upx}  
because the kinematic factor of $\kappa(m_{23}^2)$  highly breaks the $SU(3)_f$ flavor symmetry, similar to
the cases  in the  semi-leptonic charmed baryon decays.
The triangle relations derived by~\cite{Geng:2018upx,Lu:2016ogy,Gronau:2018vei} still hold 
since the isospin symmetry preserves in $\kappa(m_{23}^2)$.
% and can be recovered by the complex number parametrization.
 However, the relations from the U-spin symmetry~\cite{Grossman:2018ptn} may be broken  by $\kappa(m_{23}^2)$ 
 due to the large mass differences of hadrons.
The predicted decay branching ratio
of  ${\cal B}(\Lambda_c^+\to n \pi^{+} \bar{K}^{0})=(1.1\pm 0.1)\%$ is 3 times smaller than $(3.6\pm 0.6)\%$ 
by the BESIII observation~\cite{Ablikim:2016mcr}.
This indicates that
there exist 
some other sizable contributions to this decay, such as those from  $H(\overline{15})$, resonant states and p-wave meson pairs.
Our result for the ratio of $\frac{{\cal B}(\Xi_{c}^+\to p K^-\pi^+)}{{\cal B}(\Xi_{c}^+\to \Xi^- \pi^+\pi^+)}=0.50\pm0.13$  
is 2 times larger than the current experimental value of $0.21\pm0.04$.
For the averaged  up-down asymmetries, both $\pm\langle \alpha \rangle$ are solutions in the $\chi^2$ fitting within the $SU(3)_F$ approach,
which can be determined by experiments.
For example, one can measure
the angular distribution of the $\Lambda^0\pi^-$ pair
in the four-body decay of $\Lambda_{c}^+\to (\Xi^-\to\Lambda^0\pi^-) K^+\pi^+$  by BESIII
to fix the sign of $\langle \alpha \rangle(\Lambda_{c}^+\to \Xi^-K^+\pi^+)$, which 
has been chosen  to be negative. 
 We have also examined  the decays with the final states involving $K_{L}^{0}$/$K_{S}^0$, which contain the CF and DCS processes.
In particular, we have obtained that $\langle \alpha \rangle( \Lambda_{c}^{+} \to \Sigma^{+} K_{S}^{0} K_{L}^{0}) = -0.44 \pm 0.32 $ and $\langle \alpha \rangle( \Xi_{c}^{0} \to \Xi^{0} K_{S}^{0} K_{L}^{0}) = -0.85 ^{+0.17}_{-0.15} $, which are the same as
those for 
  the pure DCS modes of $\Lambda_{c}^{+} \to \Sigma^{+} K^{0} K^{0}$ and $ \Xi_{c}^{0} \to \Xi^{0} K^{0} K^{0} $, respectively.

%\newpage
%\section{Appendix}
% Amp1: ======================================
%\newpage

\section*{ACKNOWLEDGMENTS}
This work was supported in part by National Center for Theoretical Sciences and 
MoST (MoST-104-2112-M-007-003-MY3 and MoST-107-2119-M-007-013-MY3).

%\newpage
%\newpage

\end{document}